\documentclass [twocolumn,aps,prb,noshowpacs]{revtex4-1}
\usepackage{graphicx,amssymb,amsmath,color,bm}
\pdfoutput=1

\begin{document}
\title{Theory  of electron transport and emission from a semiconductor nanotip}
\author{Andrei Piryatinski$^1$}
\email{Corresponding author: apiryat@lanl.gov}
\homepage[ORCID:~]{https://orcid.org/0000-0001-9218-1678}
\author{Chengkun Huang$^1$} 
\homepage[ORCID:~]{https://orcid.org/0000-0002-3176-8042}
\author{Thomas J. T. Kwan$^2$}
\homepage[ORCID:~]{https://orcid.org/0000-0001-5873-7476}

\affiliation{$^1$Theoretical Division and $^2$XCP Division\\Los Alamos National Laboratory, Los Alamos, NM 87545}

\begin{abstract}
An effective mass based model accounting for the conduction band quantization in {\color{black} a high aspect ratio} semiconductor nanotip is developed to describe injected electron transport and subsequent electron emission from the nanotip. A transfer matrix formalism is used to treat electron scattering induced by the variation in the tip diameter and the electron emission. Numerical analysis of the scattering and emission probabilities is performed for the diamond parametrized nanotip model. Our scattering and emission models are further combined with a Monte Carlo (MC) approach to simulate electron transport through the nanotip. The MC simulations, also accounting for the electron-phonon scattering and externally applied electric field, are performed for {\color{black} a minimal} nanotip model and an equivalent width diamond slab. An effect of the level quantization, electron scattering due to the nanotip diameter variation, and electron-phonon scattering on the nanotip emission properties is identified and compared with the case of bulk slab.      
\end{abstract}

\date{\today}
 
\maketitle


\section{Introduction}

Nanostructured cold cathodes are required for applications in flat panel displays,\cite{ChoiAPL:1999,WangAPL:2001,Chandrasekhar:2018} flat panel X-ray sources,\cite{PosadaJAP:2014} microwave devices,\cite{Teo:2005} free-electron lasers and accelerators.\cite{liXiangkun:2013,BaryshevAPL:2014} Recent progress towards developing a new class of table top free-electron light sources and accelerators imposed a high demand in new cathode technologies.\cite{AdamoPRL:2009,PeraltaNat:2013,WongNatPhot:2015,ZhaoOptExp:2018} These accelerators are driven by infrared lasers and their dimensions are 2-3 orders of magnitude smaller than the dimensions of their radio frequency counterparts.  Therefore, high brightness cold or photo cathodes with the capability to produce low divergence and small emittance beams on the nanometer to micron scale are desirable. In light of this, nanotip based electron emission sources are natural candidates for the production of high current electron beams. Recently, carbon nanotube based field emitters have been extensively studied due to their unique charge carrier transport, emission characteristics, and well developed fabrication and characterization technologies.\cite{Bocharov:2013,Chandrasekhar:2018,ChoiAPL:1999,deHeerSci:1995,MaitiPRL:2001,Teo:2005,Bocharov:2013} Besides carbon nanotubes, the field emission has been studies in nanoparticles,\cite{DiBartolomeo:2016} nanowires (NWs),\cite{ZhiAPL:2005,Giubileo-Nmt:2017} and graphene.\cite{WuAdvMat:2009,ShaoApplSci:2018}  

Diamond naturally plays a central role in developing cold cathodes due to possibility to achieve negative electron affinity via surface hydrogenation,\cite{OkanoNat:1996,YamaguchiPRB:2009} high structural stability, chemical and radiation damage tolerance, and high thermal conductivity. A variety of nanostructured diamond field emitters have been extensively examined and fall within two major categories such as nanodiamond films and quasi-1D structures.\cite{Terranova:2015} The latter includes various assemblies of polycrystalline and/or single-crystal nanorods, NWs, nanotips, and nanopillars. Similar to carbon nanotubes an increase in the emission properties of diamond quasi-1D structures is attributed to their high aspect ratio resulting in sharp tips near which the electric field  receives significant enhancement lowering the ionization potential.\cite{Biswas:2018} Various adsorbed atoms and molecules can support resonance tunneling further enhancing the electron emission properties of carbon nanotubes and other diamond cathodes.\cite{Jarvis_JAP:2010}

In light of developing high-brightness, low emittance, fast response time, and long-lifetime electron sources, experimental study of secondary electron emission from hydrogen terminated diamond amplified photocathodes have been recently reported.\cite{BenZviBNL:2004,WangPhysRevSTAB:2011,WangPhysRevSTABexp:2011} To interpret the experiment, secondary electron generation and transport in diamond has been modeled in 3D.\cite{DimitrovJAP:2010} Subsequently, modeling of electron emission from a planar hydrogenated diamond surface accounting for associated band banding effect, effective mass anisotropy, and the inhomogeneity of electron affinity has been reported.\cite{DimitrovJAP:2015} Photocathodes of III-V semiconductors can also be activated to negative electron affinity.\cite{DowellNuclInst:2010} In accord with Spicer's 3-step photoemission model,\cite{SpicerAPA:1977} carrier photoexcitation, transport, and emission simulations, based on a Monte Carlo (MC) method, have been reported for doped GaAs bulk slab and layered structures.\cite{Karkare-MC:2013,KarkareJAP:2015Er}\cite{KarkarePRL:2014} These studies include experimental measurements on the GaAs structures with Cs/NF$_3$ surface passivation which show a good agreement with the simulation predictions. This opens opportunities for realization of ultrabright (sub)picosecond response time photocathodes based on III-V semiconductors.

Understanding field and photoemission properties of {\em nanostructured} diamond and aforementioned semiconductors requires significant modification of the models used for the bulk simulations. The modified models should take into account quantum size effects along with the electric field enhancement  near sharp geometric features on the nanoscale. An analytical model dealing with the field enhancement effect  of a nanotip has recently been reported.\cite{Biswas:2018} In this article, we address a problem of the {\em quantum size} effects on {\em electron transport} and {\em emission properties} in the case of a nanotip geometry. Details on the electron injection/photoexcitation process are not considered. However, our generic approach can be complimented by electron injection and/or photoexcitation  models resulting in a modeling tool suitable to study both electron field and photo emission processes.    

\begin{figure}[t]
\includegraphics[width=0.48\textwidth]{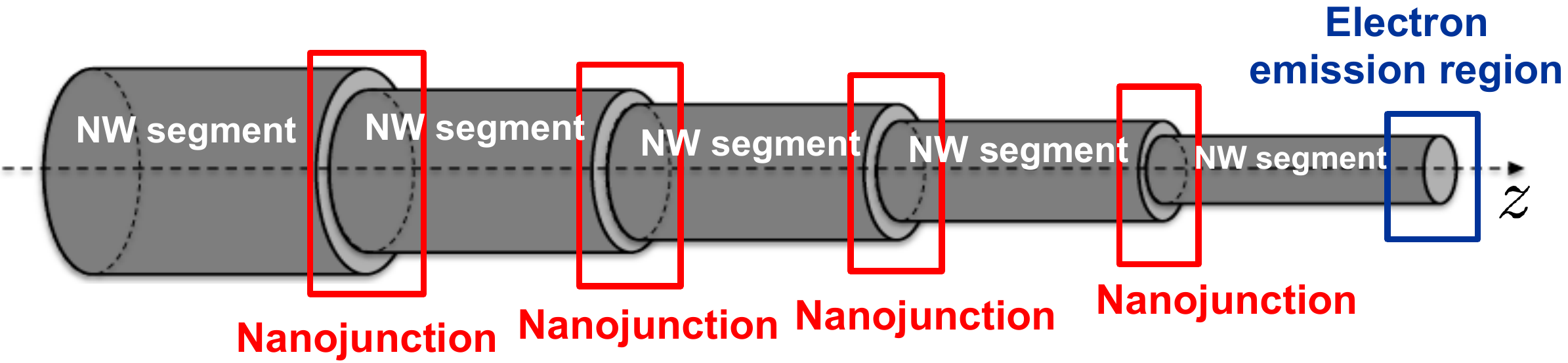}
\caption{ \color{black} Schematics of a nanotip modeled as a sequence of co-axial cylindrical NW segments of decreasing radii each supporting an incoherent electron transport. The segments are connected via nanojunctions  facilitating coherent electron scattering (Fig.~\ref{Fig-JScat}) . The last segment is terminated by the electron emission region (Fig.~\ref{Fig-NTEmis}).}   
\label{Fig-NTipMod}
\end{figure}

{\color{black} As illustrated in Fig.~\ref{Fig-NTipMod}, we model a semiconductor nanotip as a sequence of decreasing diameter co-axial cylindrical NW segments. The length of each segment is assumed to be much longer than the electron mean free path. This allows us to model the incoherent electron transport in each segment using semiclassical MC methods accounting for the electron-phonon scattering. The NW segments are separated by the nanojunctions each assumed to have a length scale smaller than the electron mean free path allowing us to neglect the electron-phonon scattering. Therefore, a {\em coherent} electron scattering occurs at each nanojunction due to the NW diameter mismatch. We also assume that the electron emission takes place only from the surface terminating the last NW segment. This assumption is valid provided  the overlap cross section of adjacent NW segments is much larger than the cross section area exposed to the vacuum preventing low energy electron emission at the nanojunctions. Such an assumption limits our model application to high aspect ratio nanotips.}

Accordingly in Sec.~\ref{Sec:ScatMdls}, a transfer matrix approach is utilized to derive the transmission and reflection coefficients (i.e., the probabilities) for the electron scattering at a nanojunction and electron emission from the surface terminating the nanotip. {\color{black} A numerical analysis of these quantities for the diamond based stand along nanojunction and a stand along emission region are provided in Sec.~\ref{Sec:ScatAnls}. Sec.~\ref{Sec:MCsmls} reports the results of the MC electron transport simulations for a minimal (two-segment) model of a diamond nanotip in the presence of an external electric field. The simulations incorporate the nanojunction electron scattering and electron emission from the surface terminating the nanotip.} Sec.~\ref{Sec:Cncl} provides concluding remarks.

\section{Theory of nanojunction scattering and nanotip emission}
\label{Sec:ScatMdls}

Let us consider a NW segment, $s$, whose symmetry axis is aligned with $z$-direction. An electron wave function in this segment is represented as a plane wave propagating in the $z$-direction within a subband designated by a set of discrete quantum numbers $\alpha$ 
\begin{eqnarray}
\label{Psi-NW-def}
    \Psi^s_{\alpha}(z) = e^{\pm ik^s_{\alpha}z}|s_\alpha\rangle.
\end{eqnarray}
Here, the electron wavevector is  
\begin{eqnarray}
\label{kzE-def}
	k^{s }_{\alpha}=\sqrt{\frac{2m_l^*}{\hbar^2}E -\frac{m^*_l}{m^*_t}\kappa^{s 2}_{\alpha}},
\end{eqnarray}
where $E$ is the total electron energy evaluated with respect to the bottom of the bulk conduction band. $m^*_l$ and $m^*_t$ are the longitudinal ($z$-direction) and the transverse (radial direction) effective masses, respectively. 

The transverse quantization of the wave vector, $\kappa^{s}_{\alpha}$, arises from the radial component of the envelope wave function represented in cylindrical coordinates, $\{\rho,\theta\}$, using Bessel functions of the first kind
\begin{eqnarray}
\label{sa-basis-def}
	\langle\rho\theta|s_{\alpha}\rangle=\frac{N^s_{\alpha}}{\sqrt{2\pi }} J_{n}(\kappa^s_{\alpha}\rho)e^{in\theta}.
\end{eqnarray}
Here, the normalization prefactor is $N^s_{\alpha} = \sqrt{2}/[\rho_sJ'_{n}(\kappa^s_{\alpha}\rho_s)]$ {\color{black} with $\rho_s$ being the radius of NW segment $s$}. Assuming infinite confinement potential on the NW surface, roots of the Bessel function $J_n(\kappa^s_{na}\rho_s)=0$, define values of the quantized radial wave vector and a set of quantum numbers $\alpha=\{n,a\}$ where $n$ is the angular momentum quantum number and $a$ is the radial quantum number.

\subsection{Nanojunction scattering model}
\label{Sec-TR}

\begin{figure}[t]
\includegraphics[width=0.45\textwidth]{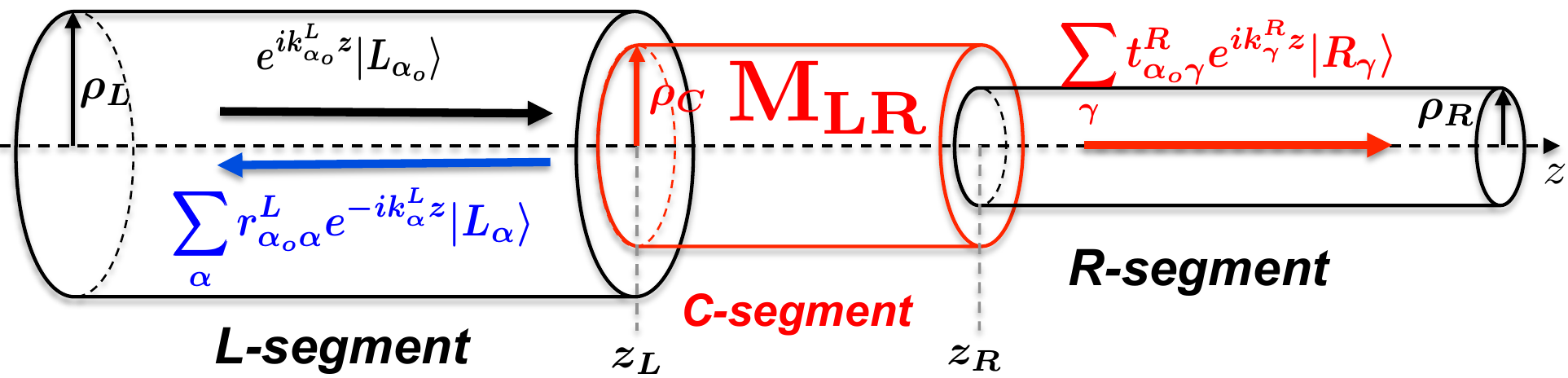}
\caption{Schematics illustrating coherent scattering of an electron from $L$ to $R$ segments forming a nanojunction via quantum scattering region $C$.  Incident (black) and reflected (blue) wave functions in $L$-segment (Eq.~(\ref{Psi-L})) are shown explicitly. Same applies to the transmitted electron wave function (red) in R-segment (Eq.~(\ref{Psi-R})). $C$-segment is described using the transfer matrix $\mathbf{M}_{LR}$ given by Eq.~(\ref{Smx-LR-1}).}   
\label{Fig-JScat}
\end{figure}

As illustrated in Fig.~\ref{Fig-JScat}, we consider an incident electron wave in the $L$-segment of radius $\rho_L$ entering a narrower ($\rho_C<\rho_L$) segment $C$. This wave further passes to a narrower segment $R$ of radius $\rho_R$ and partially reflects back to segment $L$. It is assumed that the length of the $C$-segment is $|z_R-z_L|<\xi_e$ where $\xi_e$ is electron mean free path allowing us to neglect a phonon assisted scattering. {\color{black} The finite size of the $C$-segment allows us to account for the quantum interference between the incident and reflected electron waves.}

At the boundary of segment $L$, the electron wave function, $\Psi^L_{\alpha_o}(z_L)$, is a superposition of the incident electron wave in an $\alpha_o$ subband and the electron waves reflected to subbands designated by index $\alpha$ each with an amplitude $r_{\alpha_o\alpha}$. At the $R$-segment boundary, the electron wavefunction describing transmitted electron states  $\Psi^R_{\alpha_o}(z_R)$ is a superposition of the electron waves transmitted into subbands described by $\gamma$ each with an amplitudes $t_{\alpha_o\gamma}$. These wavefunctions read  
\begin{eqnarray}
\label{Psi-L}
    \Psi^L_{\alpha_o}(z_L) &=& \sum\limits_\alpha \left( \delta_{\alpha_o\alpha} e^{ik^L_{\alpha_o}z_L}
    + r_{\alpha_o\alpha}e^{-ik^L_{\alpha }z_L}\right)|L_{\alpha}\rangle,
\\\label{Psi-R}
    \Psi^R_{\alpha_o}(z_R) &=& \sum\limits_\gamma t_{\alpha_o\gamma}e^{ik^R_{\gamma}z_R}|R_{\gamma}\rangle,
\end{eqnarray}
 where $|L_{\alpha}\rangle$ and $|R_{\gamma}\rangle$ satisfy Eq.~(\ref{sa-basis-def}) with $s=L,R$, $\delta_{\alpha_o\alpha}$ is the Kronecker delta, and the wave vectors in the exponents are given by Eq.~(\ref{kzE-def}).

As outlined in Appendix~\ref{Apx-SC}, we match the boundary conditions for the wave functions $\Psi^L_{\alpha_o}(z_L)$  and $\Psi^R_{\alpha_o}(z_R)$ with the electron wave function in segment $C$. This results in the following block-matrix equation for the reflection and transmission amplitudes   
\begin{eqnarray}
\label{Scat-LR-eq}
    \left[\begin{array}{c}\bm\Delta_{\alpha_o}\\\mathbf{r}_{\alpha_o}^L\\\end{array}\right]= 
    \begin{bmatrix} 
		\mathbf{M}^{LR}_{11} & \mathbf{M}^{LR}_{12} \\
		\mathbf{M}^{LR}_{21} & \mathbf{M}^{LR}_{22} 
	\end{bmatrix} 
    \left[\begin{array}{c}\mathbf{t}_{\alpha_o}^R\\\mathbf{0}\\\end{array}\right].
\end{eqnarray}
The column on the {\color{black} left-hand side} is a vector with the upper Kronecker $\delta$-block, $\bm\Delta_{\alpha_o}=[\delta_{\alpha_o\alpha}]$, and the lower reflection amplitude, $\mathbf{r}_{\alpha_o}^L=[r^L_{\alpha_o\alpha}]$, block. The column on the {\color{black} right-hand side} is formed from the vector $\mathbf{t}_{\alpha_o}^R=[t^R_{\alpha_o\gamma}]$ and  a column of zeros, $\mathbf{0}$, of the same size as $\mathbf{t}_{\alpha_o}^R$.  Each $\alpha$ and $\gamma$ index runs within the range of $N_{sb}$ subbands participating in the scattering.  

The $\mathbf{M}_{LR}$ transfer matrix in Eq.~(\ref{Scat-LR-eq}) contains four $N_{sb}\times N_{sb}$ blocks. According to  Appendix~\ref{Apx-SC}, the block can be evaluated from the following block matrix product    
\begin{widetext}
\begin{eqnarray}
\label{Smx-LR-1}
    \begin{bmatrix} 
		\mathbf{M}^{LR}_{11} & \mathbf{M}^{LR}_{12} \\
		\mathbf{M}^{LR}_{21} & \mathbf{M}^{LR}_{22} 
	\end{bmatrix} &=&\frac{1}{2}
    \begin{bmatrix} 
		e^{-i\mathbf{K}_Lz_L} & -ie^{-i\mathbf{K}_Lz_L}\mathbf{K}_L^{-1}\\
		e^{i\mathbf{K}_Lz_L} & ie^{i\mathbf{K}_Lz_L}\mathbf{K}_L^{-1} 
	\end{bmatrix}
    \begin{bmatrix} 
		\bm\Omega_{CL}^{-1}\mathbf{M}^C_{11}\bm\Omega_{CR} & \bm\Omega_{CL}^{-1}\mathbf{M}^C_{12}\bm\Omega_{RC}^{-1}  \\
		\bm\Omega_{LC}\mathbf{M}^C_{21}\bm\Omega_{CR} &\bm\Omega_{LC}\mathbf{M}^C_{22}\bm\Omega_{RC}^{-1}  
	\end{bmatrix}
	\begin{bmatrix} 
		e^{i\mathbf{K}_Rz_R}  & e^{-i\mathbf{K}_Rz_R} \\
		i\mathbf{K}_R e^{i\mathbf{K}_Rz_R} & -i\mathbf{K}_R e^{-i\mathbf{K}_Rz_R}
	\end{bmatrix}.
\end{eqnarray}
\end{widetext}
Here, the $C$-segment transfer matrix blocks are $\mathbf{M}^C_\text{11}=\mathbf{M}^C_\text{22}=\text{diag}[\cos\{k^C_\beta(z_{R}-z_L)\}]$,  $\mathbf{M}^C_\text{12}=\text{diag}[\sin\{k^C_\beta(z_{R}-z_L)\}/k^C_\beta]$,  and $\mathbf{M}^C_\text{21}=\text{diag}[-k^C_\beta\sin\{k^C_\beta(z_{R}-z_L)\}]$ {\color{black} with $k_\beta^C$ given by Eq.~(\ref{kzE-def}) with $s=C$}. {\color{black} Blocks} of the matrices on the right and on the left contain $\textbf{K}_L=\text{diag}[k^L_\alpha]$, $e^{i\textbf{K}_Lz_L}=\text{diag}[e^{ik^L_\alpha z_L}]$,  $\mathbf{K}_L^{-1}=\text{diag}[1/k^L_\alpha]$, and $e^{i\textbf{K}_Rz_R}=\text{diag}[e^{ik^R_\gamma z_R}]$ with the electron wavevector components defined in Eq.~(\ref{kzE-def}). Blocks $\bm\Omega_{CL}=\bm\Omega^\dag_{LC}=\left[\langle C_\beta|L_\alpha\rangle\right]$  and $\bm\Omega_{CR}=\bm\Omega_{RC}^\dag=\left[\langle C_\beta|R_\gamma\rangle\right]$ are constructed from the radial envelope wave function overlap integrals {\color{black} in which $|C_\beta\rangle$ defines radial eigenstates of segment $C$.}

According to the definition of the radial envelope function (Eq.~(\ref{sa-basis-def})), an overlap integral between two adjacent segments $s=L,C$ and $s'=C,R$ is
\begin{eqnarray}
\label{Sab}
	\Omega^{ss'}_{\alpha\beta}&=&\langle s_{\alpha}|s'_{\beta}\rangle
\\\nonumber		
        &=&N^s_{\alpha}N^{s'}_{\beta}\delta_{nn'}
		\int\limits_0^{\min\{\rho_s\rho_{s'}\}}d\rho \rho J_n(\kappa^s_{\alpha}\rho)J_{n'}(\kappa^{s'}_{\alpha'}\rho).
\end{eqnarray}
The integral vanishes for different angular momentum states ($n\neq n'$), reflecting angular momentum conservation due to the nanojunction axial symmetry. As a result,  we  consider only the scattering processes between subbands that have common quantum number $n$ but various radial numbers $a$ and $a'$.    

Equation~(\ref{Scat-LR-eq}) can be viewed as two independent sets of linear equations, $\mathbf{M}^{LR}_{11} \mathbf{t}^R_{\alpha_o}=\bm\Delta_{\alpha_o}$ and  $\mathbf{r}^L_{\alpha_o}=\mathbf{M}^{LR}_{21}\mathbf{t}^R_{\alpha_o}$. Their formal solution using matrix inversion provides the following simple representation for the transmission and reflection amplitudes in terms of the  $C$-segment transfer matrix blocks
\begin{eqnarray}
\label{tR-mx}
    \mathbf{t}_{\alpha_o}^R&=&[\mathbf{M}^{LR}_{11}]^{-1}\bm{\Delta}_{\alpha_o},
\\\label{rL-mx}    
    \mathbf{r}_{\alpha_o}^L&=&\mathbf{M}^{LR}_{21}[\mathbf{M}^{LR}_{11}]^{-1}\bm{\Delta}_{\alpha_o}.
\end{eqnarray}
  
Using transmission and reflection amplitudes, we further introduce the transmission and reflection coefficients
\begin{eqnarray}
\label{T-R-def}
    {\cal T}^{RL}_{\alpha_o\gamma}&=& \frac{{\text Re}k_{\alpha}^L} {{\text Re}k_{\alpha_o}^L} \left| t^R_{\alpha_o\gamma} \right|^2,
\\\label{R-L-def}    
   {\cal R}^{LL}_{\alpha_o\alpha}&=&  \frac{{\text Re}k_{\gamma}^R} {{\text Re}k_{\alpha_o}^L} \left| r^L_{\alpha_o\alpha} \right |^2,
\end{eqnarray}
respectively. They give us  probabilities for the electron transmission and reflection by segment $C$ and satisfy the identity 
\begin{eqnarray}
\label{TR-conserv}
    \sum_\alpha {\cal R}^{LL}_{\alpha_o\alpha}+ \sum_\gamma {\cal T}^{RL}_{\alpha_o\gamma}=1,  
\end{eqnarray}
expressing conservation of the quantum mechanical probability current.

Finally, we define expressions for the reflection and transmission coefficients describing scattering of the incident electron in a subband $\alpha_o$ of $R$-segment to $L$-segment and reflection back to segment $R$. In this case, the calculation of associated transmission, $\mathbf{t}_{\alpha_o}^L=[t^L_{\alpha_o,\gamma}]$, and reflection,  $ \mathbf{r}_{\alpha_o}^R=[r^R_{\alpha_o\alpha}]$, amplitudes should be performed according to Eqs.~(\ref{tR-mx})  and (\ref{rL-mx}) where all superscripts $R$ and $L$ are swapped, i.e., $R\rightleftarrows L$. The transfer matrix should be evaluated according to Eq.~(\ref{Smx-LR-1}) with   $R\rightleftarrows L$  including the transfer matrix argument. The transmission, ${\cal T}^{RL}_{\alpha_o\gamma}$, and reflection, ${\cal R}^{RR}_{\alpha_o\alpha}$, coefficients should be evaluated using Eqs.~(\ref{T-R-def}) and (\ref{R-L-def}) with $R\rightleftarrows L$. The current conservation condition given by  Eq.~(\ref{TR-conserv})with $R\rightleftarrows L$ should satisfy as well.

\subsection{Nanotip emission model}
\label{Sec-Emss}

\begin{figure}[t]
 \includegraphics[width=0.45\textwidth]{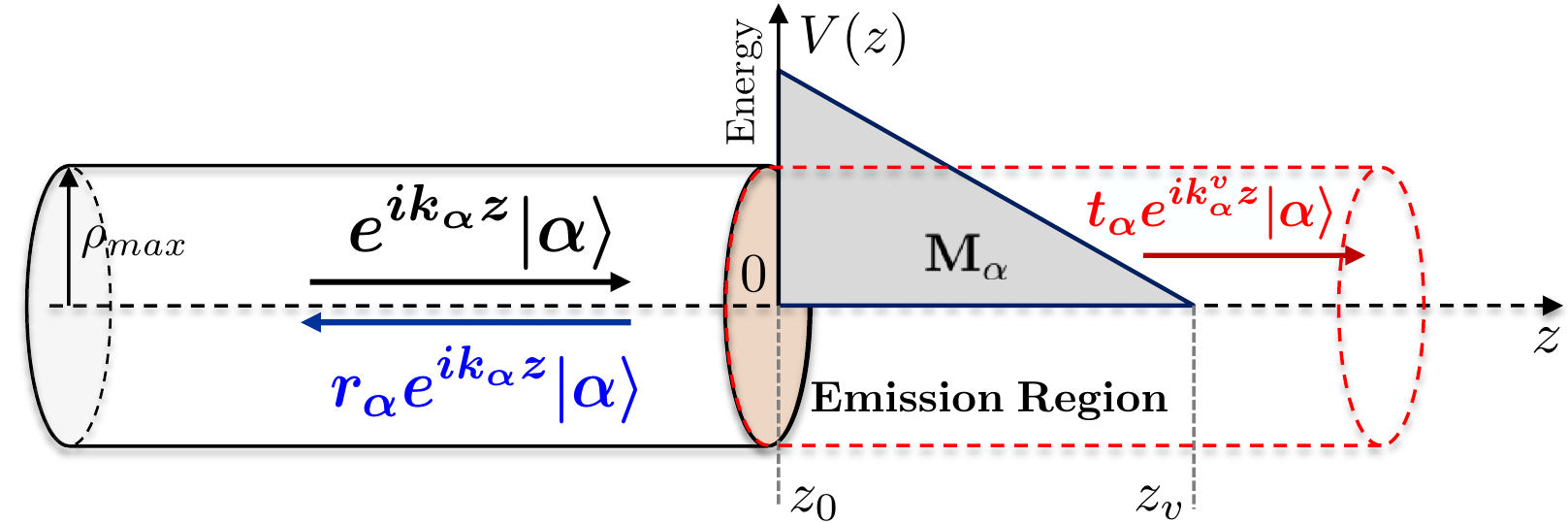}
\caption{Schematics illustrating emission region at a NW-vacuum interface.  Incident (black) and reflected (blue) wave functions in  the NW  (Eq.~(\ref{Psi-in})) are shown explicitly as well as the emitted electron wave function (red) (Eq.~(\ref{Psi-v})). The emission region, $z_0\leq z \leq z_v $, characterized by the potential barrier $V(z)$ is {\color{black} treated} with the help of the transfer matrix ${\bm M}_\alpha$ given by Eq.~(\ref{S-mx-eq}). {\color{black} Red dashed cylinder indicates the extend of radial confinement potential.}}   
\label{Fig-NTEmis}
\end{figure}

As illustrated in Fig.~\ref{Fig-NTEmis},  an electron emission is considered in the positive $z$-direction via the NW emission region extending between $z_0$ and $z_v$. The model assumes that for $z < z_0$ the surface potential is $V(z)=0$, radial confinement potential at the NW surface is infinite, and the electron effective mass is anisotropic having transverse $m_t^*$ and longitudinal $m_l^*$ components. Accordingly, the radial component of the electron envelope wave function is given by Eq.~(\ref{sa-basis-def}) where for the sake of simplicity the segment index $s$ is dropped and the angular and radial quantum numbers $\alpha=\{n, a \}$ are determined as the roots of  $J_n(\kappa_\alpha\rho_{max})=0$ with $\rho_{max}$ being the NW radius.

{\color{black} 

Once  the electron crosses  $ z_0$, its longitudinal effective mass changes abruptly to the  vacuum electron mass, $m_e$, and the surface potential becomes $V(z) > 0$ subsequently decreasing with $z$ (Fig.~\ref{Fig-NTEmis}) until reaching zero  at  $z_v$. Such a potential decrease can be associated with an external electric field applied in $z$-direction. While the electric field lowers the surface potential $V(z)$, it is reasonable to assume that it has no effect on {\em the radial} confinement potential at the interval $z_0\leq z\leq z_v$ and keep it infinite. Such an assumption allows us to describe  the electron envelope wave function using Eq.~(\ref{sa-basis-def}) with the set of quantum numbers $\alpha$ at the interval $z_0<z < z_v$. The confinement potential can further be set to zero at some point $z_v\lesssim z$ and electron diffraction outside this region introduced. 
}

To find electron emission probability, the scattering theory approach is applied. The NW electron wave function $ \Psi_\alpha(z_0)$ at the entrence to the emission region, $z_0$, is a superposition of the incident and reflected plane electron  waves. Each wave is weighted with the confined radial state $|\alpha \rangle$ satisfying Eq.~(\ref{sa-basis-def}).  {\color{black}The  electron wave function $ \Psi^v_\alpha(z_v)$ exiting the emission region at $z_v$ is also a plane wave weighted with the same envelope wave function $|\alpha \rangle$ due to the presence of the radial confinement potential.} Accordingly,
\begin{eqnarray}
\label{Psi-in}
    \Psi_\alpha(z_0) &=&e^{ik_{\alpha }z_0}|\alpha \rangle + r_{\alpha}e^{-ik_{\alpha}z_0}|\alpha \rangle,
\\\label{Psi-v}
    \Psi^v_\alpha(z_v) &=& t_\alpha e^{ik^v_{\alpha }z_v}|\alpha \rangle,
\end{eqnarray}
with $r_\alpha$ and $t_\alpha$ being the emission region reflection and transmission amplitudes, respectively. The longitudinal electron wavevector, $k_\alpha$, is given by Eq.~(\ref{kzE-def}) with the superscript $s$ dropped for the sake of simpilisity. The emitted electron longitudinal wavevector reads
\begin{eqnarray}
\label{kzEv-def}
	k^v_{\alpha}=\sqrt{\frac{2m_e}{\hbar^2}E-\kappa_{\alpha}^2}.
\end{eqnarray}
Notice that Eq.~\eqref{kzEv-def} depends on the electron mass in vacuum, $m_e$, and the total electron energy, $E$. Since the energy is conserved during the scattering process, it can be evaluated from Eq.~(\ref{kzE-def}) as 
\begin{eqnarray}
\label{EofEz}
	E=\frac{\hbar^2k_\alpha^2}{2m_l}+\frac{\hbar^2\kappa_{\alpha}^2}{2m_t^*}.
\end{eqnarray}

Appendix~\ref{Apx-STnl} outlines some details on matching the boundary conditions at $z_0$ and $z_v$ for the wavefunctions in Eqs.~(\ref{Psi-in}) and (\ref{Psi-v}) using the transfer matrix formalism. This calculation results in the following matrix equation 
\begin{eqnarray}
\label{ST-mx-eq}
    \left[\begin{array}{c}1\\r_{\alpha}\\\end{array}\right]= 
    \begin{bmatrix} 
		M^\alpha_{11} & M^\alpha_{12} \\
		M^\alpha_{21} & M^\alpha_{22} 
	\end{bmatrix} 
    \left[\begin{array}{c}t_{\alpha}\\0\\\end{array}\right],
\end{eqnarray}
for the reflection and transmission amplitudes at $z_0$ and $z_v$, respectively. The $2\times 2$ transfer matrix describing the emission region is
\begin{widetext}
\begin{eqnarray}
\label{S-mx-eq}
     \begin{bmatrix} 
		M^\alpha_{11} & M^\alpha_{12} \\
		M^\alpha_{21} & M^\alpha_{22} 
	\end{bmatrix}  =\frac{1}{2}
    \begin{bmatrix} 
		e^{-ik_{\alpha}z_0}  & -i(m^*_l/m_e) e^{-ik_{\alpha}z_0}/k_{\alpha} \\
		e^{ik_{\alpha}z_0}  & i (m^*_l/m_e) e^{ik_{\alpha}z_0} /k_{\alpha} 
	\end{bmatrix} 
    \begin{bmatrix} 
		{\cal M}^\alpha_{11} & {\cal M}^\alpha_{12} \\
		{\cal M}^\alpha_{21} & {\cal M}^\alpha_{22}
	\end{bmatrix}
    \begin{bmatrix} 
		e^{ik^v_{\alpha}z_v}  & e^{-ik_{\alpha}z_v}  \\
		ik^v_{\alpha}e^{ik^v_{\alpha}z_v}  & -ik^v_{\alpha} e^{-ik^v_{\alpha}z_v} 
	\end{bmatrix},
\end{eqnarray}
and depends on the transfer matrix 
\begin{eqnarray}
\label{MT-tmx}
    \begin{bmatrix} 
		{\cal M}^\alpha_{11} & {\cal M}^\alpha_{12}\\
		{\cal M}^\alpha_{21} & {\cal M}^\alpha_{22}
	\end{bmatrix}  
    =\prod_{j}  
    \begin{bmatrix} 
		\cos[k^j_{\alpha}(z_{j}-z_{j+1})] & \sin[k^j_\alpha(z_{j}-z_{j+1})]/k^j_{\alpha}\\
		-k^j_\alpha\sin[k^j_{\alpha}(z_{j}-z_{j+1})] & \cos[k^j_{\alpha}(z_{j}-z_{j+1})]
	\end{bmatrix},
\end{eqnarray}
\end{widetext}
associated with the surface potential $V(z)$. Here, the matrix product runs over small intervals along $z$-direction numerated by index $j$. Within each of these intervals, the surface potential is approximated by a rectangular potential barrier of a height $V(z_j)$. The  longitudinal momentum entering Eq.~(\ref{MT-tmx}) is
\begin{eqnarray}
\label{kzEj-def}
	k^j_{\alpha}=\sqrt{\frac{2m_e}{\hbar^2}\left[E-V(z_j)\right] -\kappa_{\alpha}^2},
\end{eqnarray}
with the total energy $E$ provided in Eq.~(\ref{EofEz}). 

Solving Eq.~(\ref{ST-mx-eq}) for the transmission and reflection amplitudes 
\begin{eqnarray}
\label{t-tun}
    t_{\alpha} &=& 1/M^\alpha_{11},
\\\label{r-tun}
    r_{\alpha} &=& M^\alpha_{21}/M^\alpha_{11},   
\end{eqnarray}
we further introduce the transmission and reflection coefficients
\begin{eqnarray}
\label{Ttun-def}
    {\cal T}_{\alpha}&=& \frac{k^v_{\alpha}} {{\text Re}k_{\alpha}} \frac{m_l^*} {m_e}\left| t_{\alpha} \right|^2,
\\\label{Rtun-def}    
   {\cal R}_{\alpha}&=&  \left| r_{\alpha} \right |^2,
\end{eqnarray}
describing electron emission and reflection probabilities. They satisfy the quantum mechanical current conservation condition  
\begin{eqnarray}
\label{TR-tun-cons}
    {\cal R}_{\alpha}+{\cal T}_{\alpha}=1.  
\end{eqnarray}


\section{Analysis of scattering and emission processes}
\label{Sec:ScatAnls}

{\color{black} This section presents numerical analysis of the electron scattering  at a stand along nanojunction (Fig.~\ref{Fig-JScat}) and the electron emission from a stand along NW segment (Fig.~\ref{Fig-NTEmis}). The NWs are assumed to be made of diamond crystalline structures with the $[100]$ valley aligned along  $z$-direction. The nanojunction is set to have the radii $\rho_L=3$~nm, $\rho_C=2.5$~nm, and $\rho_R=2$~nm and the lengths of $C$-segment $z_R-z_L=0.5$~nm. We further chose the radius of  NW segment participating in the electron emission to be $\rho_{max}=\rho_R=2$~nm to set the stage for the simulations discussed in Sec.~\ref{Sec:MCsmls}.}

{\color{black} The effective mass in the $z$-direction is identified as longitudinal and set to the bulk value $m_l^*=1.4m_e$. In the transverse direction, the effective mass is also set to its bulk value $m_t^*=0.36m_e$.\cite{DimitrovJAP:2015} Notice that for the adopted radii of the NW segments, the anisotropy and nonparabolicity of the bulk conduction band can affect the values of the quantized subband effective masses.\cite{NeophytouIEEE:2008} For the proof of principle calculations discussed below it is enough to use the bulk values. However, precise device modeling will require to account for the size quantization corrections to the bulk effective masses.}

A convenient shorthand notation $\alpha$ was introduced in Sec.~\ref{Sec:ScatMdls} to denote a set of angular and radial quantum numbers $\{n,a\}$ that designate quantized electron subbands in NW segments. Below, specific values of these quantum numbers will be considered. Therefore, all expressions containing these quantities will show their explicit rather than the shorthand representation. 

\begin{figure}[t]
 	\includegraphics[width=0.4\textwidth]{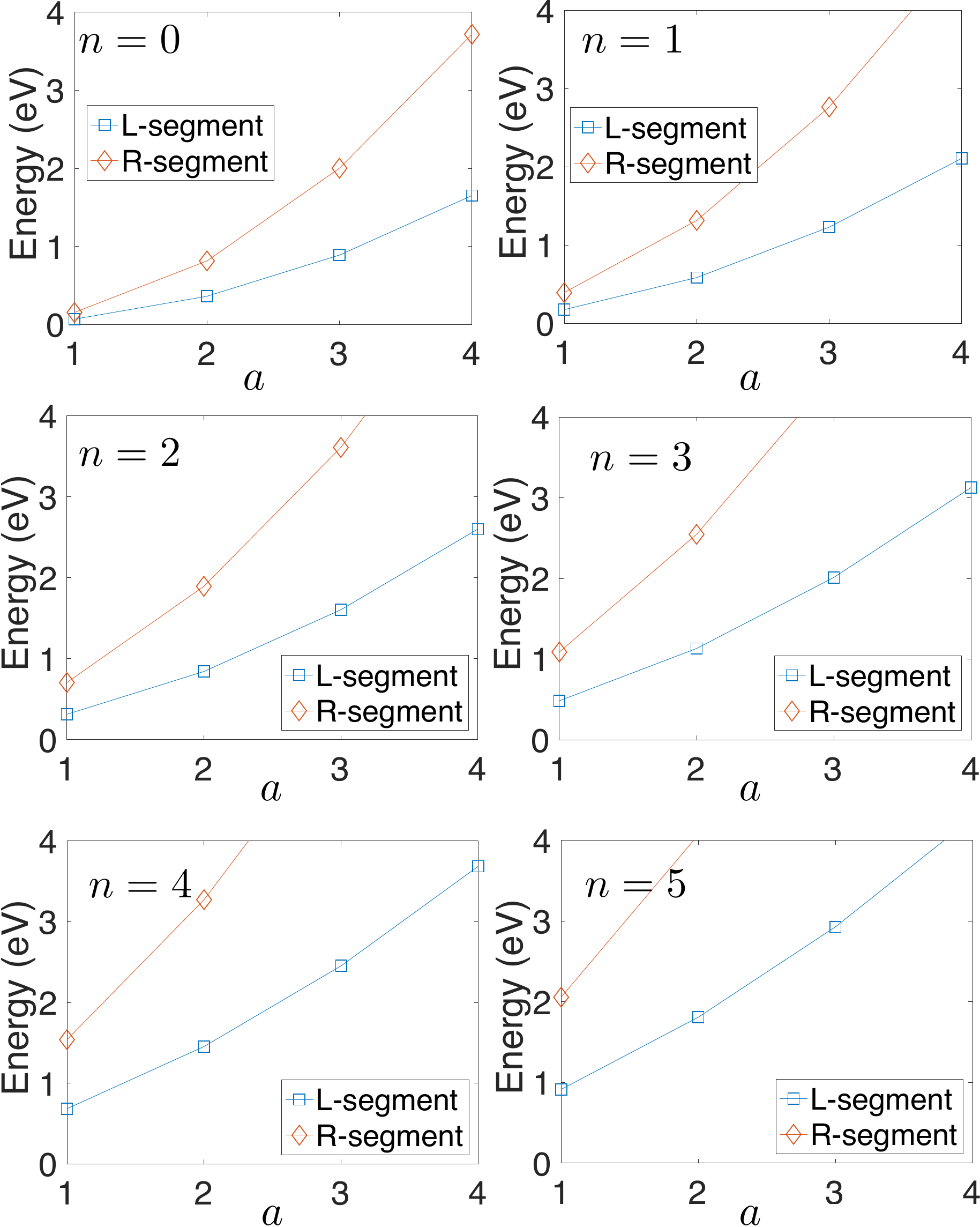}
\caption{Quantized energy values for the minima of conduction electron subbands as a function of radial quantum number $a$ calculated for the $L$ and $R$ diamond NW segments. Each panel presents the energy spectrum for a fixed angular momentum quantum number that varies from $n=1$ to $5$. Solid lines are given as a guide for the eye.}   
\label{Fig-Ena}    
\end{figure}

We start with Fig.~\ref{Fig-Ena} presenting calculated values of the quantized energy minima, ${\hbar^2\kappa_{n a}^2}/{2m_t^*}$, associated with conduction subbands of the $L$ and $R$ segments. The plot indicates a rapid (fractions of eV) growths of the energy values as the radial quantum number $a$ increases. The rapid growth is attributed to a small value, $m_t^*=0.36m_e$, of the transverse effective mass.  Since $\rho_L>\rho_R$, the energy values in segment $R$ are always larger than those in segment $L$ given the same quantum numbers $n$ and $a$. 

\subsection{Nanojunction scattering}
\label{Sec:Scatnj}

The transmission and reflection coefficients for the incident electron placed into a subband  $\{n_o, a_o\}$ of either $L$ or $R$ segment given kinetic energy $E_z=\hbar^2 {k^s_{n_o a_o}}^2/2m^*_l$ ($s=L,R$) are calculated using the formalism developed in Sec.~\ref{Sec-TR}. Keeping in mind that the scattering conserves the angular momentum, we set $n=n_o$ for the transmitted and reflected electron states.

\begin{figure}[t]
 	\includegraphics[width=0.4\textwidth]{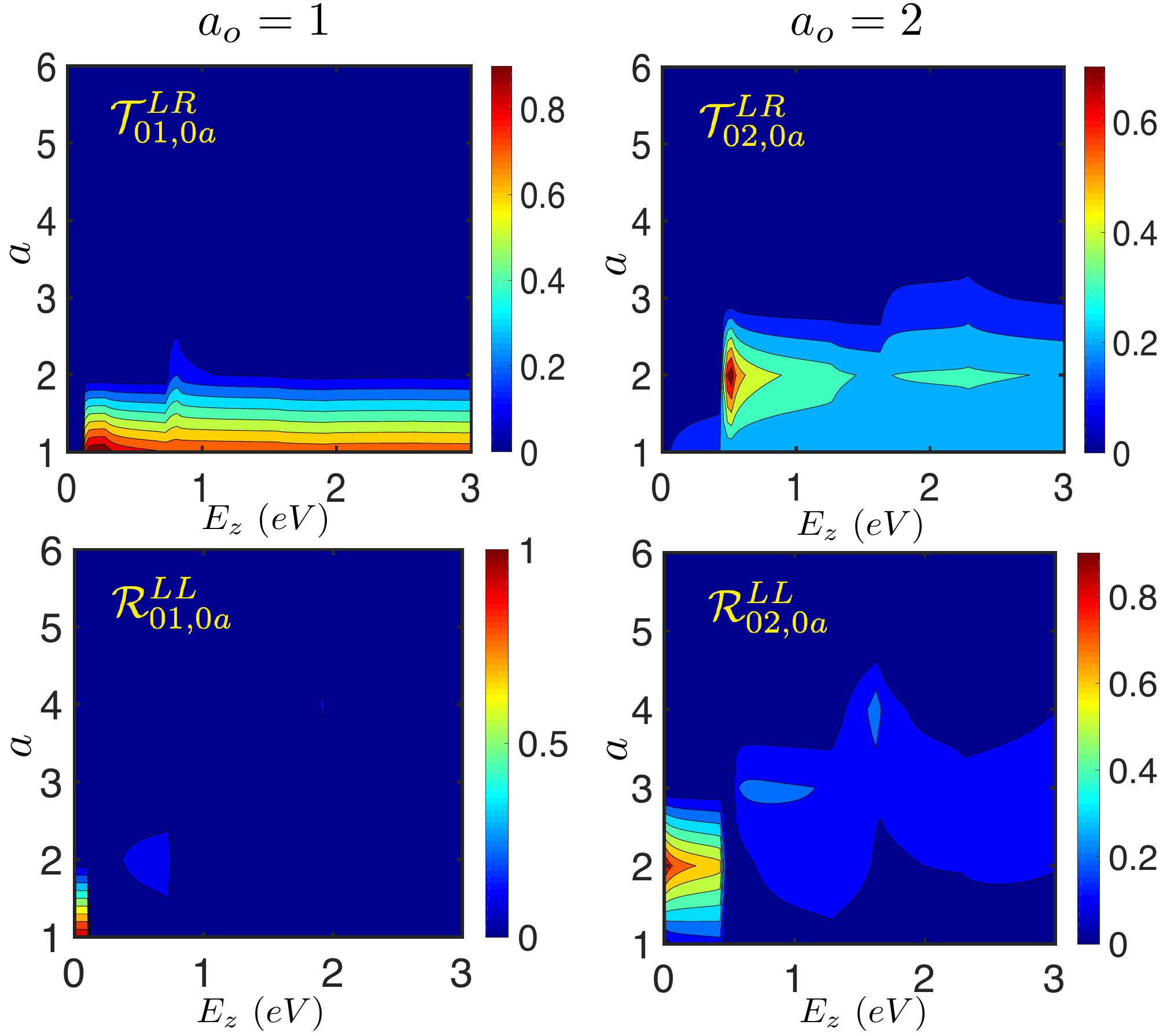}
\caption{Contour plot of  $L$ to $R$ transmission coefficient ${\cal T}_{n_oa_o,n_oa}^{LR}$ and the $L$ to $L$ reflection coefficient ${\cal R}_{n_oa_o,n_oa}^{LL}$ calculated for the diamond NW junction shown in Fig.~\ref{Fig-JScat}. The incident electron is placed into an $a_o$ subband with a kinetic energy $E_z$ and scatters to subbands characterized by $a$. The angular quantum number is set to $n_o=0$ in all panels. Left (right) column shows the results for $a_o=1$ ($a_o=2$).}
\label{Fig-TRLR}   
\end{figure}
The calculated transmission coefficient (i.e., the probability), of an incident electron within $n_o=0$ and $a_o=1$ ($n_o=0$ and $a_o=2$) subband of $L$-segment to be scattered to $a$ subbands of $R$-segment as a function of the kinetic energy $E_z$ is presented in the left (right) top panel of Fig.~\ref{Fig-TRLR}. Associated reflection coefficients are given in the lower row of Fig.~\ref{Fig-TRLR}. Provided an incident electron is in the subband $a_o=1$, the left column of Fig.~\ref{Fig-TRLR} indicates the following trends: For the electron kinetic energy within the range of  $0<E_z<86$~meV, the electron gets reflected back into the state $a_o$. Once the kinetic energy passes a threshold of $86$~meV an efficient transmission opens into the subband $a=1$ of segment $R$. According to Fig.~\ref{Fig-Ena}, the activation energy threshold can be identified as a gap between $a_o=1$ subband in segment $L$ and $a=1$ subband in segment $R$. Thus, observed suppression of electron transmission below the energy threshold can be associated with the electron energy conservation.  

\begin{figure}[t]
 	\includegraphics[width=0.4\textwidth]{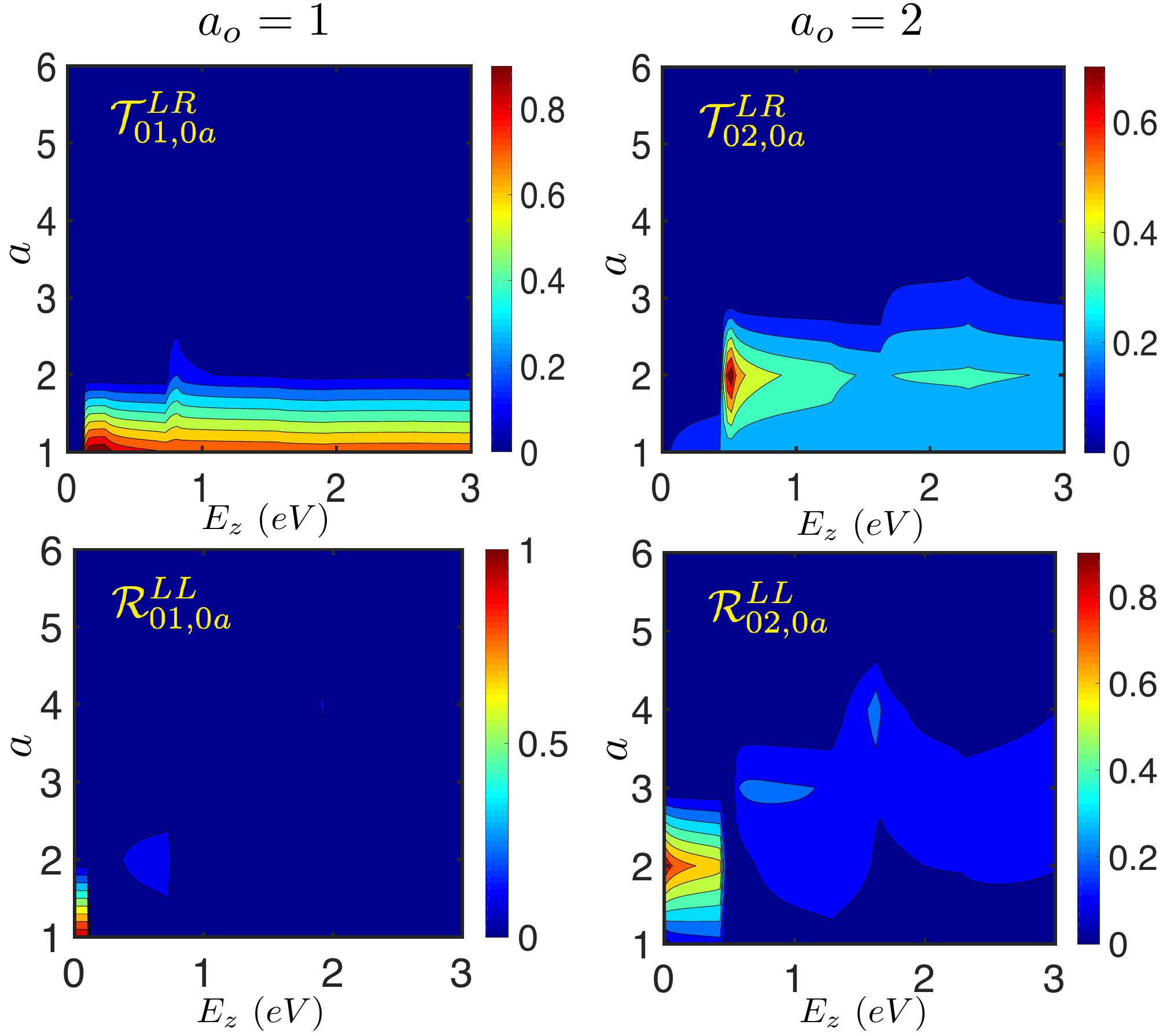}
\caption{ The same as in Fig.~\ref{Fig-TRLR} but an incident electron is in $R$-segment and gets transmitted to $L$ segment with probability ${\cal T}_{0a_o,0a}^{RL}$ and reflected back with probability ${\cal R}_{0a_o,0a}^{RR}$.}   
\label{Fig-TRRL} 
\end{figure}

The right panel in Fig.~\ref{Fig-TRLR} describes the electron scattering from the subband $a_o=2$. The activation energy gap for the transmission to $a=2$ subband of $E_z=450$~meV clearly shows up in the plot. Besides  strong reflection back within the $a_o$ subband, a weak transmission to the $a=1$ subband of $R$-segment is observed for $E_z<450$~meV. According to Fig.~\ref{Fig-Ena}, the $a=1$ $L$-subband is lower in energy than the $a_o=2$ $R$-subband and, thus,  does not have any activation threshold. Observed low values of the transmission coefficient are due to small values of associated  overlap integral (Eq.~(\ref{Sab})). The same explains lack of an efficient reflection to $a=1$ subband. Furthermore, vanishing overlap integrals between the subbands separated by larger energy gaps completely suppress  the scattering to those bands as seen in Fig.~\ref{Fig-TRLR}  for $E_z$ ranging up to $3$~eV. Similar trends are observed for the transmission and reflection amplitudes of the incident electron placed into $R$-segment as can be seen by comparing Figs.~\ref{Fig-TRLR} and \ref{Fig-TRRL}.

Finally, we point out that slow modulation of the transmission and reflection coefficients with increasing $E_z$ can be attributed to the effect of quantum interference {\color{black} associated with the finite size of segment $C$}.  The electron wavelength scales as $\lambda_z\sim E_z^{-1/2}$. Our estimate shows that the values of $E_z =0.1$~eV, 1.0~eV, and 3.0~eV corresponds to $\lambda_z=3.3$~nm, 1.0~nm, and 0.6~nm, respectively. Taking into account that the length of the $C$-segment is set to 0.5~nm, the variation of the kinetic energy in the range from 0~eV to 1.0~eV results in the electron wavelength values exceeding the region size and making the interference effect negligible. The variation of energy kinetic between 1~eV and 3~eV results in the wave length changes from twice to single C-segment length. In this case a slow modulation of the transmission and reflection probabilities occurs as observed in the plot. 

To summarize, our analysis shows that, for the adopted parameters, efficient transmission through the NW junction occurs between subbands which are very close in energy. An efficient reflection typically happens for small values of kinetic energy. For electron kinetic energy of hundreds meV and below, the quantum interference has weak effect on the transmission and reflection coefficients provided the size of the interference region is about the electron mean free path. The quantum interference becomes important if $E_z\gtrsim 1$~eV.

\subsection{Nanotip emission}
\label{Sec:Emis}

Using the formalism developed in Sec.~\ref{Sec-Emss}, we model electron emission from the NW segment terminating a nanotip. The surface potential is  adopted to have the form
\begin{eqnarray}
\label{Vt-def}
    V(z)=\chi- F(z-z_0),
\end{eqnarray}
for $z>z_0$ with $\chi$ being the electron affinity and $F$ being the amplitude of an external electric field applied along $z$-axis. For the diamond surface terminating the NW at $z_0$ we set $\chi=0.3$~eV.\cite{DimitrovJAP:2015} The external electric field values are varied within the range of $F=5$~MeV to 20~MeV. For these values, the length of the emission region, $z_v-z_0=\chi/F$, varies between 60~nm and 15~nm, respectively. 

\begin{figure}[t]
 	\includegraphics[width=0.45\textwidth]{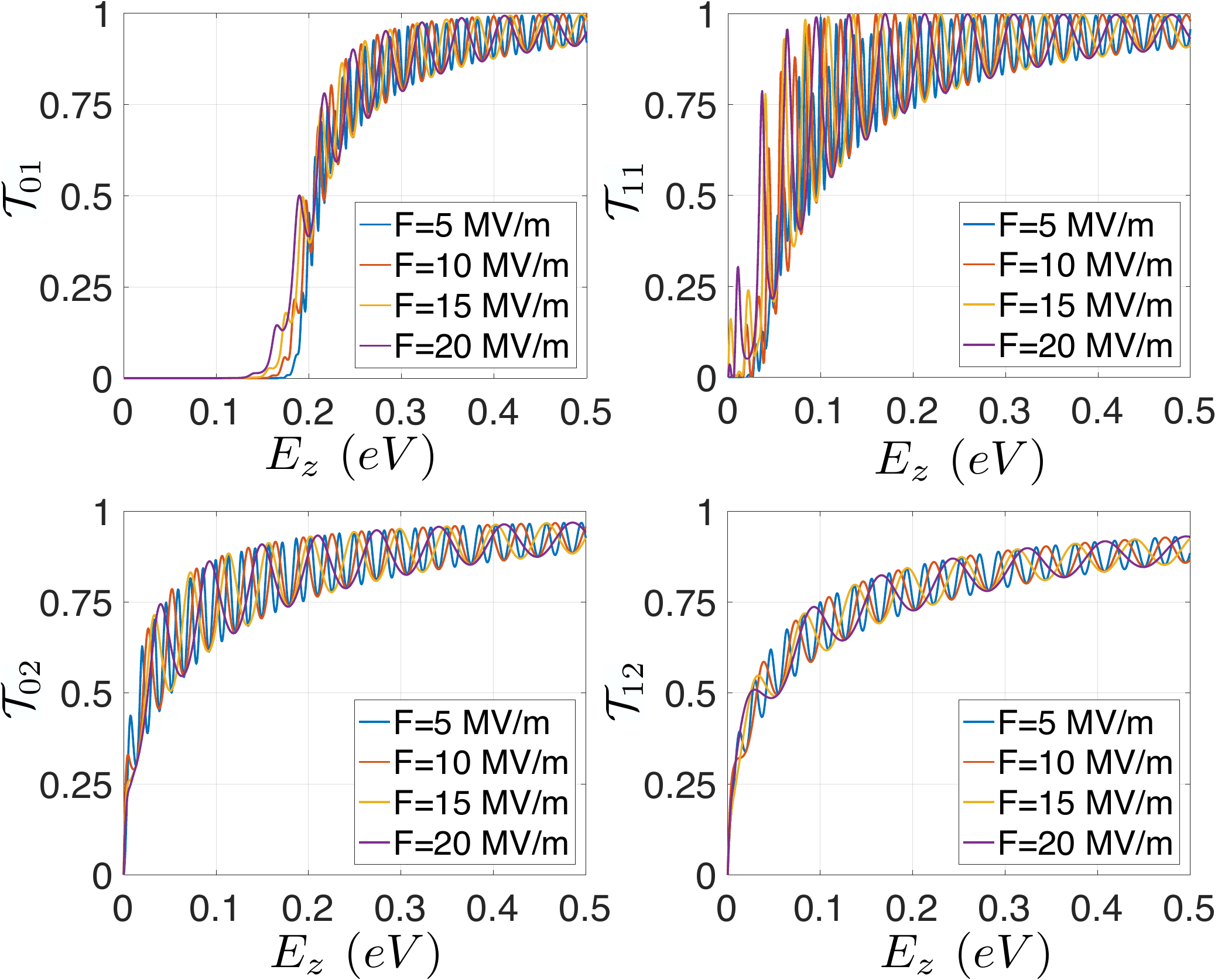}
\caption{Calculated transmission coefficient ${\cal T}_{na}$ for the electron emission as a function of its kinetic energy $E_z$ in the presence of an external electric field $F$ with the values specified in the insets.}   
 \label{Fig-Tems}
\end{figure}

Left column of Fig.~\ref{Fig-Tems} shows  calculated transmission coefficients for the electron emission from subbands characterized by the angular quantum number $n=0$ and the radial quantum numbers $a=1,2$. In the case of ${\cal T}_{01}$, the curves pick up fast oscillatory behavior around $E_z=0.2$~eV indicating that the electron experiences scattering at the very top of the potential barrier.  Taking into account that  the electron affinity is $\chi=0.3$~eV,  a source of the energy shift $\Delta E_{01}=\chi-E_z\sim 0.1$~eV allowing the electron to clear the surface potential barrier needs clarification.  

The energy dependance of transmission (Eq.~(\ref{Ttun-def})) and reflection (Eq.~(\ref{Rtun-def})) coefficients within the emission region is determined by transfer matrix in the form of Eqs.~(\ref{S-mx-eq})--(\ref{MT-tmx}) that depends on the electron wavevector, $k^j_{na}$, given by Eq.~(\ref{kzEj-def}).  Taking into account that the energy is conserved during the emission, we substitute Eq.~(\ref{EofEz}) into Eq.~(\ref{kzEj-def}). This results in
{\color{black}
\begin{eqnarray}
\label{kzEj-meff}
	k^j_{na}=\sqrt{\frac{2m_e}{\hbar^2}\left(E_z+\Delta E_{na}-V(z_j)\right)},
\end{eqnarray}
where  $E_z$ denotes the incident electron kinetic energy  within $\{n,a\}$ subband, $V(z_j)$ is the surface potential value at specified coordinate, and 
\begin{eqnarray}
\label{DEa-def}
\Delta E_{na}=\left[1-\frac{m_t^*}{m_e}\right]\frac{\hbar^2\kappa_{na}^2}{2 m_t^*}
\end{eqnarray}
is the energy shift due to the quantized band energy resulting from the electron mass mismatch at the interface. The effect of the effective mass mismatch has already been examined for the bulk-vacuum interface in relation with the transverse momentum conservation.\cite{DimitrovJAP:2015} Below, we examine new consequences of this effect rising fomr the band quantization. 
}

If the wave vector (Eq.~(\ref{kzEj-meff})) is imaginary, the electron tunnels under the barrier. Once it becomes real, the electron clears top of the barrier and the transmission coefficient acquires  an oscillatory behavior. According to Eq.~(\ref{kzEj-meff}), either real or imaginary value of the wavevector depends on whether the incident electron kinetic energy, $E_z$, is larger or smaller than the {\em effective} potential energy barrier $V(z_j)-\Delta E_{na}$. According to Eq.~(\ref{DEa-def}), the contribution of $\Delta E_{na}$ to the effective potential can be zero if $m_l^*=m_e$. If $m_l^*<m_e$ or $m_l^*>m_e$ then $\Delta E_{na}$ acquires either positive sign (decreasing the potential barrier) or negative sign (increasing the potential barrier). 

{\color{black} In the adopted case of diamond NW, $m_t^*=0.36m_e$ resulting in {\em positive} values of $\Delta E_{na}$. Specifically, $\Delta E_{na}$=0.1~eV shifts the emission energy threshold to $E_z=200$~meV as observed in Fig.~\ref{Fig-Tems} for ${\cal T}_{01}$. For the case of ${\cal T}_{11}$, the quantized energy contribution is even higher, $\Delta E_{11}=0.25$~eV, further shifting the barrier crossing threshold to $\sim 50$~meV. The values of $\Delta E_{02}=0.52$~eV and  $\Delta E_{12}=0.84$~eV exceed the electron affinity, $\chi=0.3$~eV resulting in the fast oscillatory rise of the  ${\cal T}_{02}$ and ${\cal T}_{12}$ starting at $E_z=0$ as clearly seen in the lower panels of Fig.~\ref{Fig-Tems}. }

\section{Electron transport and emission simulations}
\label{Sec:MCsmls}

{\color{black} In this section, we discuss the results of our electron transport and emission simulations performed for a minimal nanotip model. Following generic representation of a nanotip in Fig.~\ref{Fig-NTipMod}, we narrow down the model to {\em two} NW segments, namely $L$ and $R$ connected via a single nanojunction.  Adopted $L$ and $R$ NW segment radii are defined in Sec.~\ref{Sec:ScatAnls} as $\rho_L=3.0$~nm and $\rho_R=2.0$~nm, respectively. The length of each segment is set to $z_L-z_i=z_0-z_R=100$~nm, where $z_i$ ($z_0$) is the coordinate of the left (right) end of $L$ ($R$) NW segment. In our minimal model, the electrons are emitted from the right end of segment $R$ characterized by coordinate $z_0$.}

To perform electron transport simulations, a bulk MC device simulation code\cite{VasileskaMC:2010} was modified to run 1D trajectories within quantized subbands. Following Ref.~[\onlinecite{Ramayya:2008}], we implemented a model for electron scattering by the acoustic and optical phonons in NWs. This model along with the bulk electron-phonon scattering model is briefly summarized in Appendix~\ref{Apx-Scph}.  In all simulations discussed below, electrons with the kinetic energy satisfying the Boltzmann distribution at  temperature of 300~K are injected into a single subband at the left end of segment $L$ (at $z_i$). {\color{black} An external electric field $F$ = 20~MV/m is applied along $z$-axis. The field value is set larger than a typical field applied in an experiment to partially account for the geometric field enhancement effect. Since we focus on the quantum-size effects, the space charge effect is not included in the simulations.} Using bulk value of the diamond DC dielectric constant $\epsilon=5.7$, the field inside is estimated to be $F_{NW}=3.57$~MV/m. Trajectory  run time is chosen such that all electrons are either emitted or reflected back to the injection point.

\begin{figure}[t]
 	\includegraphics[width=0.48\textwidth]{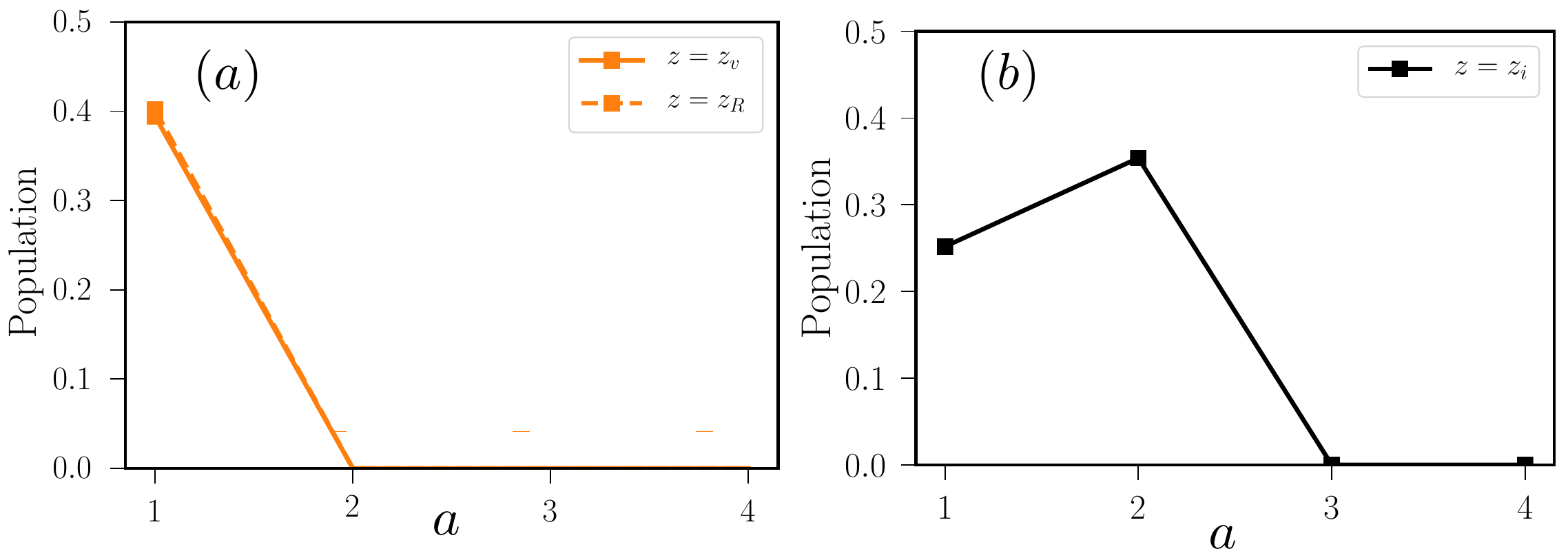}
\caption{Normalized subband population distribution for electrons (a) transmitted through the junction (dashed line), emitted to the vacuum (solid line), and (b) reflected back to initial point $z_i$ during 3~ps transport simulation with the phonon scattering turned off. Electrons are initially injected into $n=1$ and $a=2$ subband. The scattering does not change quantum number $n$. Lines are a guide for the eye.}   
 \label{Fig-n1a2-pp}
\end{figure}

Electron quantum scattering at the $L$-$R$ junction and the electron emission and reflection at $z_0$ are accounted for via transmission and reflection coefficients introduced above with the same parameters as used in Sec.~\ref{Sec:ScatAnls}.  An electron reaching the $L$-$R$ nanojunction during the MC simulation can cross the junction during a free-flight step. In this case, the transmission and reflection probabilities are cumulated and renormalized for all possible scatterings across the junction into various outgoing quantum states. A random number is generated and compared to the range of the cumulated probability corresponding to each possible scattering selecting the outgoing  quantum states. The electron is then either transmitted or reflected, i.e., placed either at $z_R$ or $z_L$ coordinate depending on the scattering direction. Its kinetic energy is updated to conserve the total energy. For the case of electron emission and reflection at the right end of $R$-segment, the similar algorithm is applied.

\subsection{Transport without electron-phonon scattering}
\label{Sec:TranspPhoff}

\begin{figure}[t]
 	\includegraphics[width=0.35\textwidth]{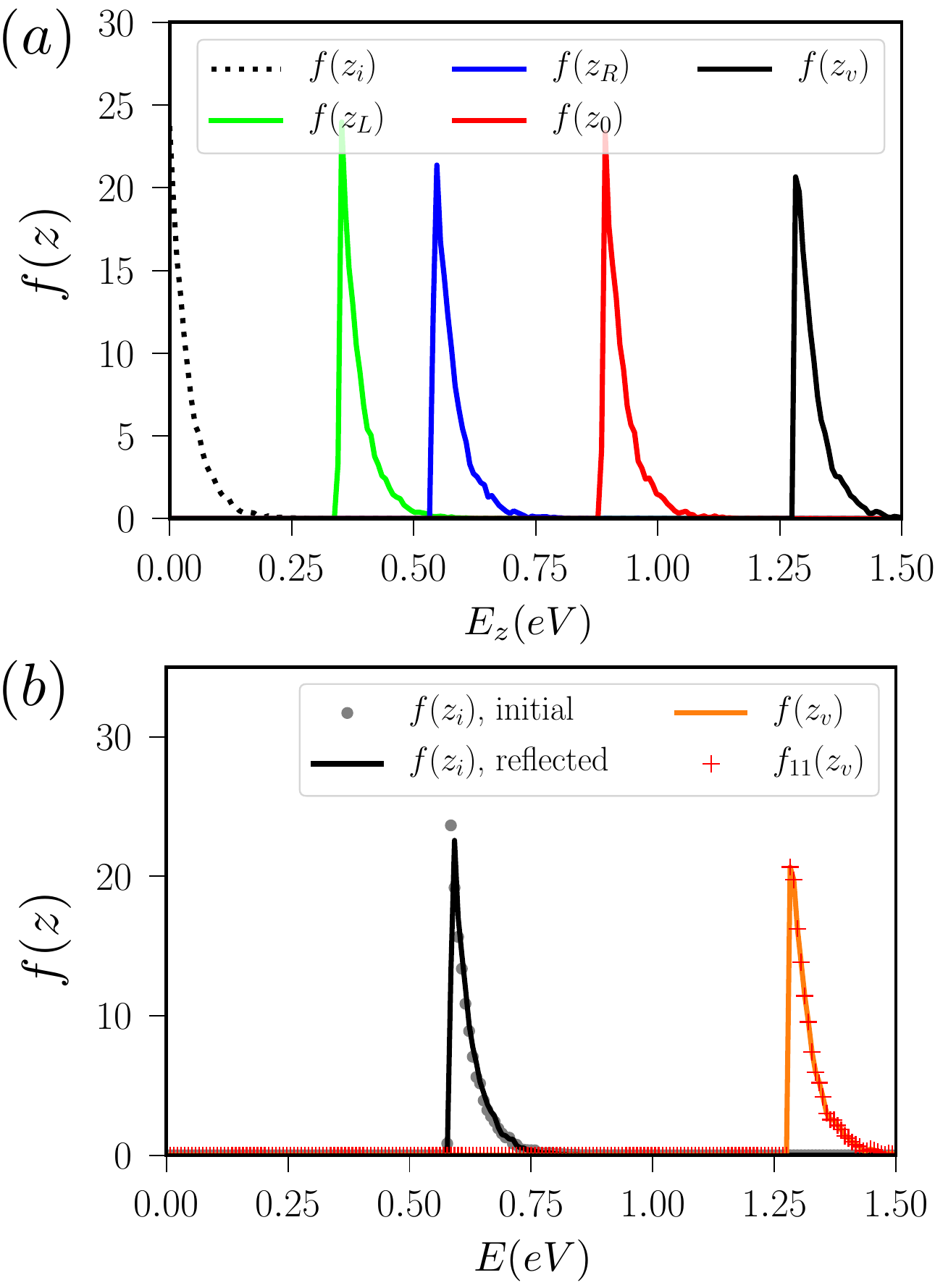}
\caption{ (a) Kinetic energy, $E_z$, distribution for the electrons moving forward sampled at the injection point $z_i$, at the nanojunction boundaries $z_L$ and $z_R$, and at the emission region boundaries $z_0$ and $z_v$. (b) Total energy, $E$, distribution for the injected, reflected, and emitted electrons. Here, $f(z)$ denotes the electron distribution function summed over all populated subbands whereas $f_{11}(z)$ stands for the distribution function associated with the $n=1$, $a=1$ subband. The same electron trajectories as in Fig.~\ref{Fig-n1a2-pp} are used to obtain the energy distributions.}   
 \label{Fig-n1a2-f}
\end{figure}

To clarify the effect of the nanojunction and the emission region scattering, results of the simulations performed with the phonon scattering turned off are discussed first.  The electrons are initially injected into the $n=1$ and $a=2$ subband whose bottom energy $\hbar^2\kappa_{12}^2/2m^*_t\simeq 0.58$~eV (Fig.~\ref{Fig-Ena}).  Subsequently, $3$~ps trajectories are run and the electron population and energy distributions are recorded at different sampling points.

The subband population distribution normalized per total number of injected electrons sampled at the exits of the nanojunction, $z_R$, and  the emission region, $z_v$, as well as at the injection coordinate, $z_i$, are given in  Fig.~\ref{Fig-n1a2-pp}. Remember that the angular momentum conservation during the scattering processes  preserves the quantum number $n=1$. Panel~(a) shows that during the course of simulation $40$\% of injected electrons (dashed line) are transmitted through the junction into a single $n=1$, $a=1$ subband of NW segment $R$ and {\color{black} subsequently emitted to vacuum}. According to panel (b), electrons reflected back to the injection point, $z_i$, populate the subbands $a=1,2$. This observation is in agreement with the discussion in Sec.~\ref{Sec:ScatAnls} identifying that the junction scattering occurs within a narrow subband distribution. 

\begin{figure}[t]
 	\includegraphics[width=0.48\textwidth]{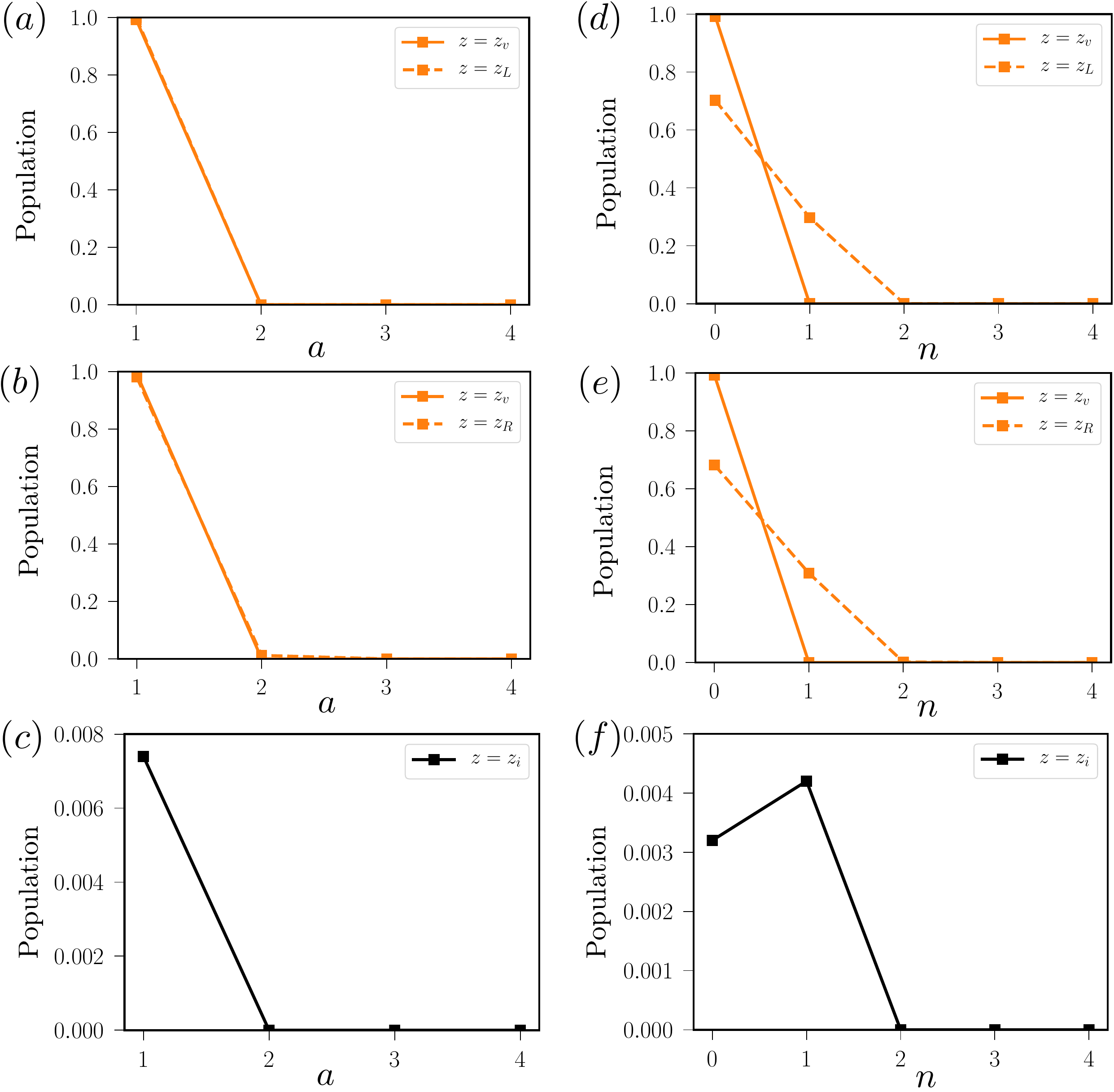}
\caption{Normalized subband population distribution for the electrons (a) approached the nanojunction (dashed line) and emitted to vacuum (solid line), (b) transmitted through the junction (dashed line) and emitted (solid line) to vacuum, and (c) reflected back to initial point $z_i$ during 4~ps transport simulation accounting for the phonon scattering. Electrons are initially injected into the $n=1$ and $a=2$ subband. In (a)--(c) the population is summed over the angular quantum number $n$. Panels (d)--(f) provide population distributions at the same sampling points as in (a)--(c), respectively, but as a function of $n$  and are summed over the radial quantum number $a$.  Lines are a guide for the eye.}   
 \label{Fig-n1a2-ph-pp}
\end{figure}

Distributions of the {\em kinetic energy}, $E_{z}$, for the forward moving electrons at different points of the nanotip are shown in Fig.~\ref{Fig-n1a2-f}~(a).  Each distribution has the form of Boltzmann function and is energy upshifted {\color{black} from} the initial distribution, $f(z_i)$. The shape of each distribution is preserved because of the vanishing dependance of the transmission coefficients on the energy within the distribution width and lack of the electron-phonon scattering. The kinetic energy shift of $0.35$~eV between distributions at the injection point, $z_i$, and the nanojunction entrance, $z_L$, as well as between the point of the nanojunction exit, $z_R$, and the emission region entrance, $z_0$, reflects electrons acceleration by the external  electric field.   

The kinetic energy difference of $\sim 0.2$~eV for the distributions describing electrons entering, $f(z_L)$, and exiting, $f(z_R)$, the $L$-$R$-junction is the energy  difference between the $n=1$, $a=2$ subband in $L$-segment and the $n=1$, $a=1$ subband of $R$-segment. For the electrons entering, $f(z_0)$, and exiting, $f(z_v)$,  the emission region, the shift in the kinetic energy is determined by the momentum variation provided by Eqs.~\eqref{kzEv-def} and \eqref{EofEz}.  {\color{black} Notice that the electron kinetic energy distribution at the entrance to the emission region, $f(z_0)$, (red) peaks $\sim 0.9$~eV which is above the potential barrier ($\chi=0.3$~eV). Accordingly, all electrons entering the emission region clear the barrier as can be seen in Fig.~\ref{Fig-n1a2-pp}a.} 
 
\begin{figure}[t]
 	\includegraphics[width=0.35\textwidth]{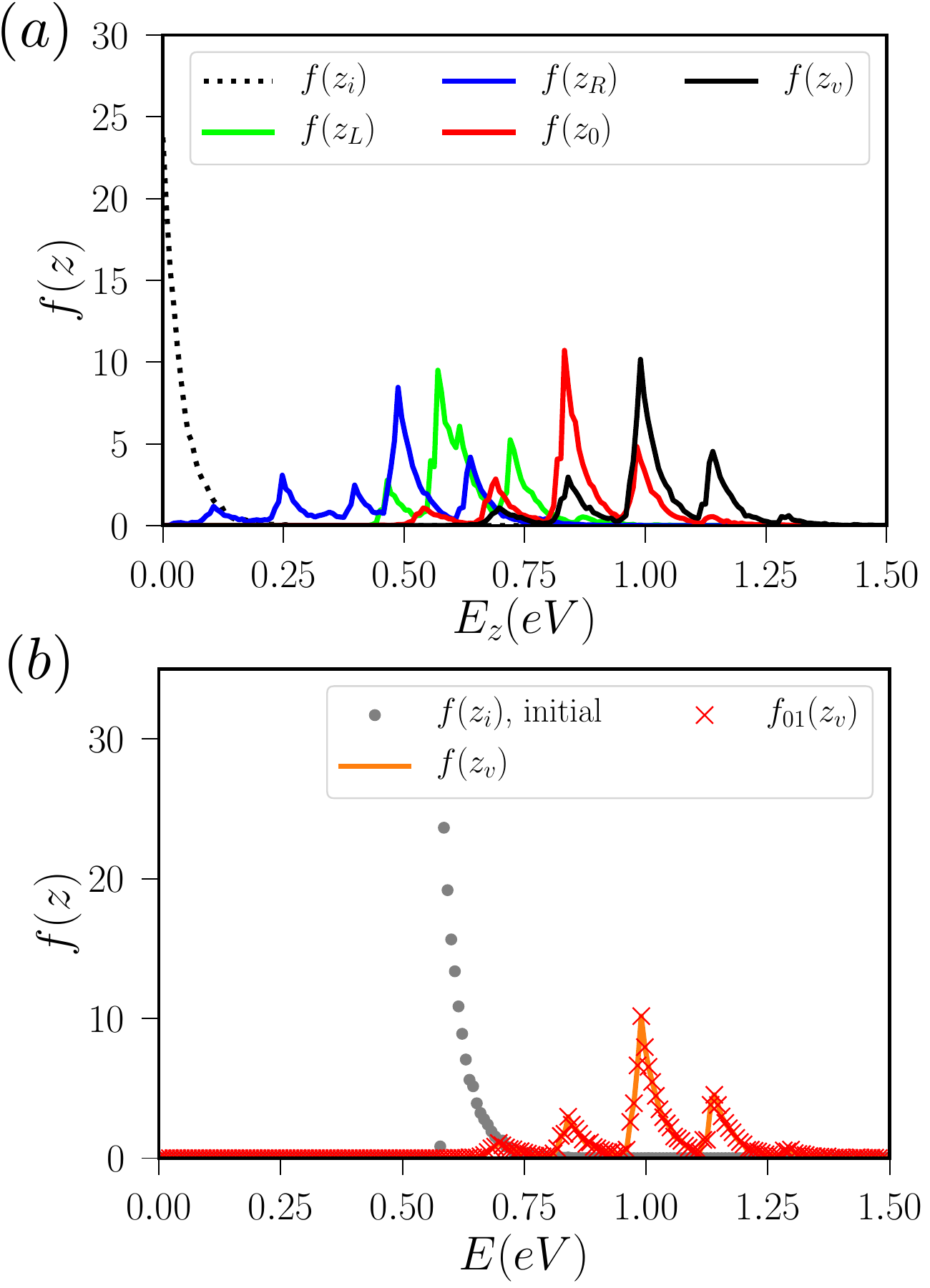}
\caption{ (a) Kinetic energy, $E_z$, distribution for the electrons moving forward sampled at the injection point $z_i$, at the nanojunction boundaries $z_L$ and $z_R$, and at the emission region boundaries $z_0$ and $z_v$. (b) Total energy, $E$, distribution for the injected and emitted electrons.  No reflected electron energy distribution is shown due to very small number of the reflected electrons [Fig.~\ref{Fig-n1a2-ph-pp}~(c) and (f)]. Here, $f(z)$ denotes the electron distribution function summed over all populated subbands whereas $f_{01}(z)$ stands for the distribution function associated with the $n=0$, $a=1$ subband. The same electrons trajectories as in Fig.~\ref{Fig-n1a2-ph-pp} are used to obtain the energy distributions.   }   
 \label{Fig-n1a2-ph-f}
\end{figure}

The {\em total energy}, $E$, distributions for the electrons  emitted to  vacuum and reflected back are shown in Fig.~\ref{Fig-n1a2-f}~(b). Specifically, the energy distribution for all emitted electrons (solid orange line) is compared with the distribution of electrons emitted from $n=1$, $a=1$ subband (orange crosses). The distributions coincide which is in agreement with Fig.~\ref{Fig-n1a2-pp}~(a) showing that the emission occurs solely from  $n=1$, $a=1$ subband. Although the electrons scattered back are distributed between $n=1$, $a=1,2$ subbands (Fig.~\ref{Fig-n1a2-pp}~(b)), the coincidence between the total energy distribution for all electrons scattered back (solid black line) and their initial distribution (grey dots) in Fig.~\ref{Fig-n1a2-f}~(b) reflects conservation of the total energy (i.e., the absence of the dissipation processes). The upshift of $0.7$~eV between the  emitted and injected/reflected energy distribution is due to the electron acceleration by the external electric field.

\subsection{Transport with electron-phonon scattering}
\label{Sec:TranspPhon}

Given the same, $n=1$, $a=2$, initial conditions, we now turn the electron-phonon scattering on and examine associated electron trajectories which were run on the timescale of $4$~ps. The normalized electron population distributions are presented in Fig.~\ref{Fig-n1a2-ph-pp}.  Since the electron-phonon scattering processes do not conserve electron angular momentum, the population distributions are presented as functions of the radial and angular quantum numbers. Panels (a) and (d) clearly show that during the transport through $L$-segment, the phonons cause population relaxation down to subbands $n=1$, $a=1$ and $n=0$, $a=1$. According to panels (b) and (e), electron transmission through the nanojunction occurs to the same subbands. In particular $n$ is preserved due to the momentum conservation by the nanojunction scattering. While moving through $R$-segment, electrons further relax to the lowest energy band ($n=0$, $a=1$) and $\sim 99$\% of the injected electrons get emitted. According to panels (c) and (f) only $\sim 0.7$\% of the electrons are reflected back to the injection point by the nanojunction.

\begin{figure}[t]
 	\includegraphics[width=0.48\textwidth]{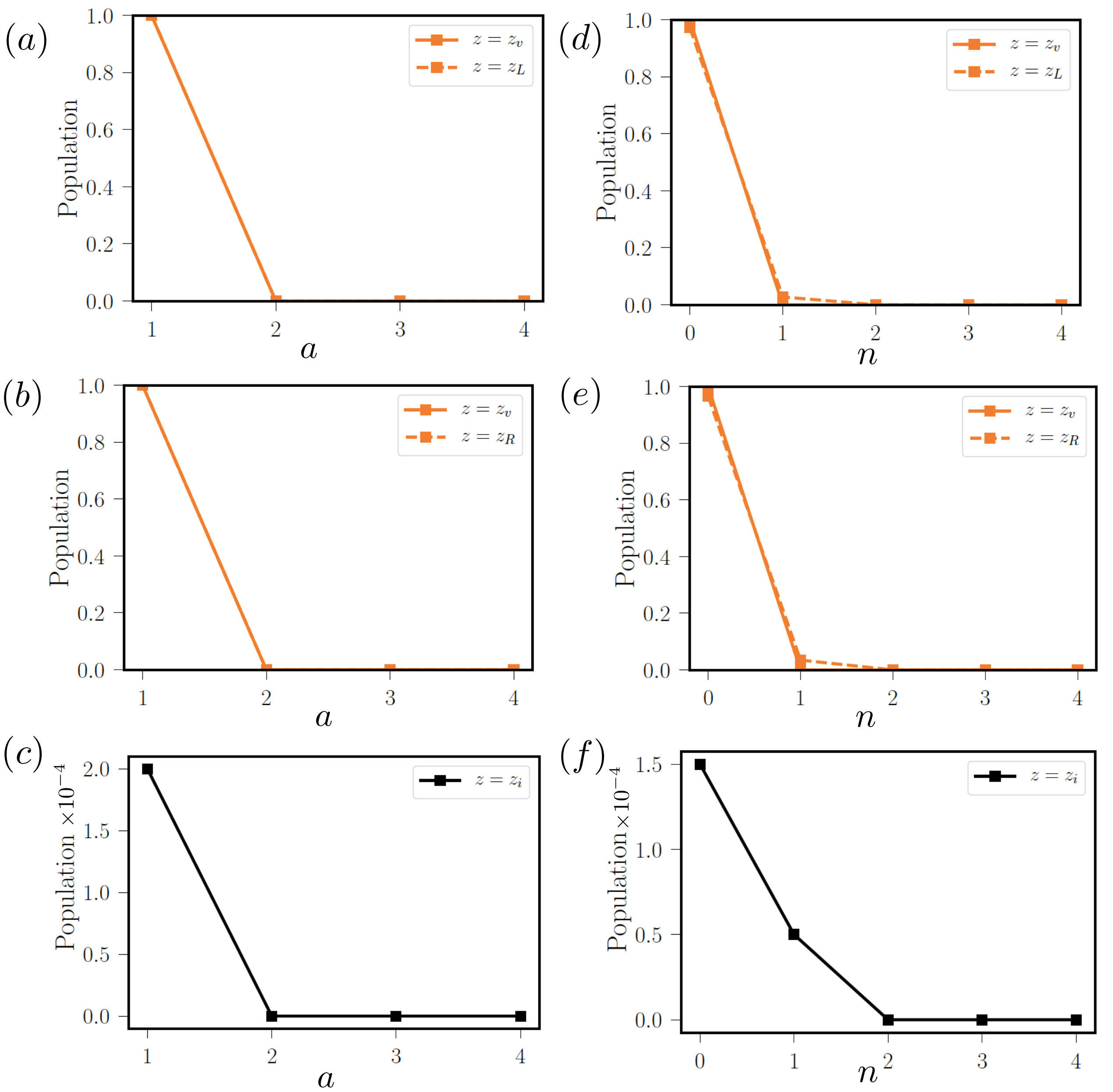}
\caption{Same as in Fig.~\ref{Fig-n1a2-ph-pp} but the electrons are initially injected into the $n=0$ and $a=1$ subband and the trajectories are run for 5~ps.}   
 \label{Fig-n0a1-ph-pp}
\end{figure}

Comparison of the kinetic and the total energy distributions in the case of phonon scattering turned off (Fig.~\ref{Fig-n1a2-f}) and on (Fig.~\ref{Fig-n1a2-ph-f}), reveals that the latter distribution functions are significantly broadened due to the inter- and intraband population transfer already identified in Fig.~\ref{Fig-n1a2-ph-pp}. The side peaks appear as the phonon replicas of the main distribution peaks shifted up and down for the energy of the optical phonon quantum. In Fig.~\ref{Fig-n1a2-ph-f}~(b), the coincidence between the energy distribution function of the electrons {\em emitted from the subband} $n=0$, $a=1$, $f_{01}(z_v)$,  with the distribution of {\em all} electrons emitted by the nanotip, $f(z_v)$, is also in agreement with  Fig.~\ref{Fig-n1a2-ph-pp}, confirming that the emission occurs from the lowest energy band. Notice that, in Fig.~\ref{Fig-n1a2-ph-f}~(a), the kinetic energy of the electrons entering the emission region, $f(z_0)$, (red) is distributed above $0.5$~eV. This is above the electron affinity level ($\chi=0.3$~eV) and according to Fig.~\ref{Fig-Tems}  results in transmission probability  one.

\begin{figure}[t]
 	\includegraphics[width=0.35\textwidth]{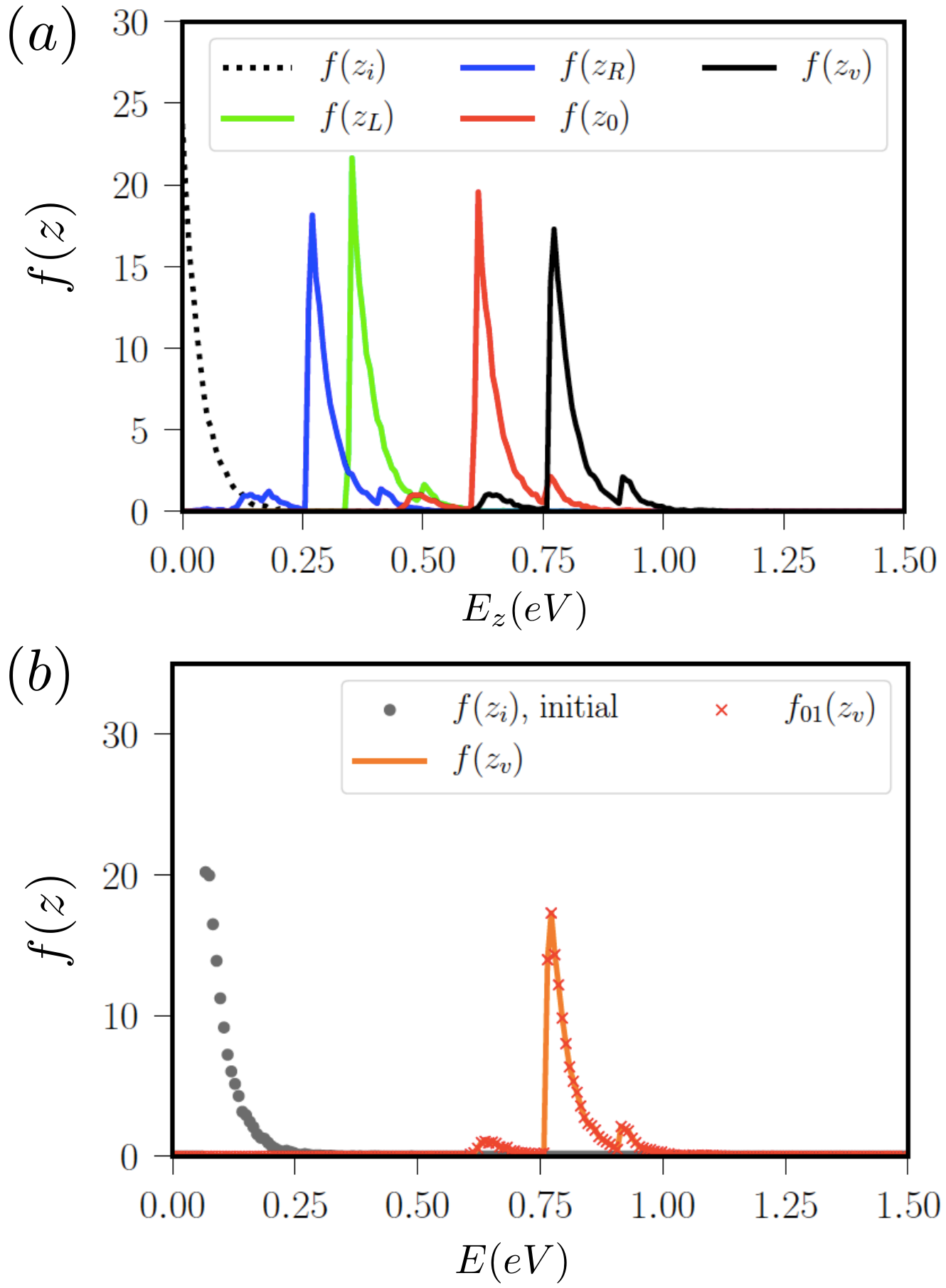}
\caption{ Same as in Fig.~\ref{Fig-n1a2-ph-f} but the electrons are initially injected into the $n=0$ and $a=1$ subband, and the trajectories are run for 5~ps.}   
 \label{Fig-n0a1-ph-f}
\end{figure}

\subsection{Comparison of nanotip and bulk emission}
\label{Sec:TranspBlk}

To compare the nanotip electron transport and emission properties with those in bulk diamond, we injected the electrons into the lowest energy subband ($n=0$, $a=1$) of the nanotip.  Furthermore, a bulk slab of equivalent, 200~nm, thickness and equivalent crystallographic axes orientation was considered. Electrons were injected at one side of the slab to the bottom of [100] valley forming the same Maxwell distribution as in the case of the nanotip and further accelerated by the same external electric field.  The MC device simulation code\cite{VasileskaMC:2010} was used without any modifications adopting a bulk phonon scattering model summarized in Appendix~\ref{Apx-Scph}. The emission from the slab was modeled using the same potential as for the nanotip. In both cases the electron trajectories were run for $5$~ps. 

According to Fig.~\ref{Fig-n0a1-ph-pp}~(a), (b), (d), and (e), the electron transport is confined to the lowest energy subband of both $L$ and $R$ NW segments. About 100\% of the injected electrons are emitted. According to panels~(c) and (f) of the same figure, a negligible amount, 0.02\%, of the injected electrons is reflected back. The  energy distributions in Fig.~\ref{Fig-n0a1-ph-f} are weakly influenced by the phonon assisted scattering. The phonon assisted scattering rate in a NW depends on the density of final states that scales as $\sim 1/\sqrt{E^z_{na}}$ representing the van Hove singularity  [Eqs.~\eqref{W1Dac} and \eqref{W1Dop}]. $E^z_{na}$ denotes kinetic energy of the scattered electrons. Accordingly, the sharply peaked density of states allows the scattering only between or near the bottoms of the subbands. In our case, the interband scattering is suppressed, since electrons do not gain enough  kinetic energy to be scattered to the higher energy subbands. Observed high emission rate has the same origin as in the case of Sec.~\ref{Sec:TranspPhon}.

\begin{figure}[t]
 	\includegraphics[width=0.35\textwidth]{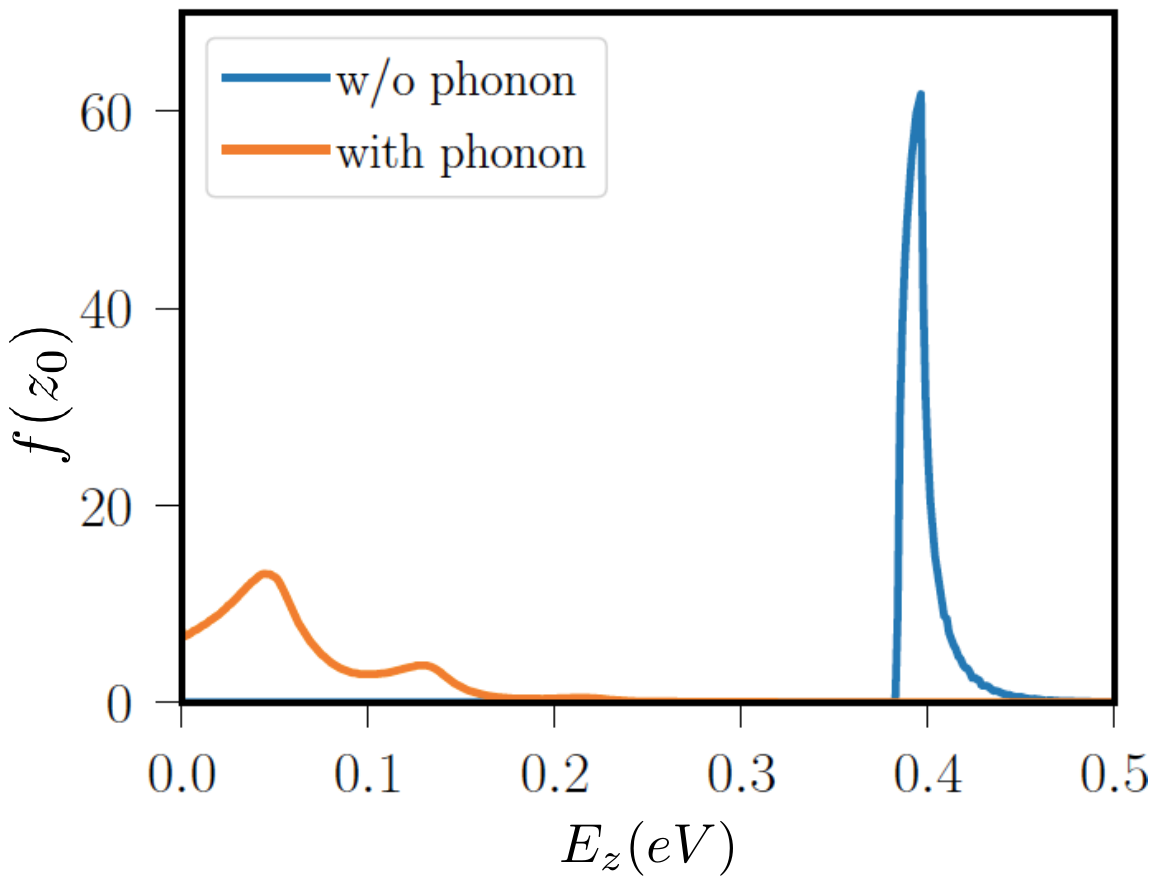}
\caption{  Kinetic energy, $E_z$, distribution for the electrons reaching  the emission region at $z_0$ for a bulk slab calculated with and without phonon scattering.}   
 \label{Fig-bulk-f}
\end{figure}

In the case of bulk, 100\% of electrons reach the emission region but only 21\% of them are emitted. This effect can be rationalized by looking at the kinetic energy distribution of the electrons, $f(z_0)$, reaching the emission region presented in Fig.~\ref{Fig-bulk-f}. According to the plot, the phonon assisted scattering has profound effect on the  transport by reducing the electron energy to the range below $0.2$~eV. This is below the electron affinity value of 0.3~eV. Thus, the electrons have to tunnel under the barrier to escape to the vacuum which significantly lowers the emission rate. Compared to the nanotip case, high phonon scattering rate originates from the scaling of the scattered electron density of states which is $\sim \sqrt{E}$ [Eqs.~\eqref{W3Dac} and \eqref{Wopiv-bulk}]. In contrast to the sharply peaked van Hove singularity in NWs, bulk square root dependance allows for a broad range of states accessible by the scattered electrons. 

\section{Concluding remarks}
\label{Sec:Cncl}

We have developed a theory addressing electron scattering by diameter variation in a semiconductor nanotip and electron emission from the surface terminating the nanotip. The theory is applied to examine transmission and reflection probabilities for the varying diameter nanojunction and for the emission region of a diamond nanotip. The analysis shows that the nanojunction scattering occurs within a narrow range of quantized conduction electron subbands due to the fast decrease of the envelope function overlap. As further demonstrated, the mismatch in the NW transverse effective mass and electron mass in vacuum within the emission region along with the band energy quantization can result in a substantial lowering of the emission potential and the enhancement of the electron emission from the nanotip. Noteworthy that the emission surface passivation should affect the emission properties. Such effects cannot be accounted directly in the effective mass approximation and are subject of separate studies utilizing, e.g., atomistic surface modeling. {\color{black} Atomistic modeling of the emission region can also be employed to validate adopted distribution of the confinement and the emission surface potentials (Fig.~\ref{Fig-NTEmis}) and make necessary model refinements. Ultimate validation of the proposed emission model requires comparison with experimental measurements.}

The scattering and emission models have been implemented into the MC transport simulations and used to simulate the transport and emission properties of a minimal geometry nanotip. The simulations show that the interplay of the electron-phonon scattering and the quantum size effects can result in up to 100\% emission of the injected electrons. In contrast, the phonon assisted scattering in an equivalent size and orientation bulk slab of diamond leads to a significantly lower emission outcome. This emphasizes an advantage of nanostructuring to achieve an efficient electron source. {\color{black} Noteworthy, that the nanotip surface roughness and electron-impurity scattering effects are not accounted for in the model. Such effects can potentially lower desired electron emission efficiency and require further examination.}  Although the simulations are performed for the case of a diamond based nanostructure, our model is general and can be parametrized to study the same effects in nanotips of various semiconductor materials, e.g., GaAs. Finally, the methodology sets the stage for investigation of the field and/or  photoemission dynamics in semiconductor nanotips subject to implementation of specific electron injection or photoexcitation models.

\acknowledgements
This work is supported by the Laboratory Research and Development (LDRD) program at Los Alamos National Laboratory. A. P. would like to acknowledge David H. Dunlop for stimulating discussions of the scattering and emission models. We would like to thank Evgenya I. Simakov and Bo Kyoung Choi for comments on the manuscript. 

\vspace{20pt}
{\bf Author contributions:} A.P., C.H., T.J.T.K. conceived and developed the core ideas. A.P. developed and implemented the nanoscale scattering models. C.H. implemented and performed the MC transport and emission simulations. All authors took part in the result analysis. A.P. and C.H. wrote this manuscript.

\appendix

\section{Derivation of $\textbf{M}_{LR}$ transfer matrix}
\label{Apx-SC}

Conduction band electron wavefunction in the $C$-segment is defined as
\begin{eqnarray}
\label{Psi-C}
    \Psi^C(z) &=& \sum\limits_\beta \phi_\beta(z)|C_\beta\rangle,
\end{eqnarray}
where $\phi_\beta(z)$ is a superposition of counter propagating electron waves characterized by the wavevectors $\pm k_\beta^C$ both defined by Eq.~(\ref{kzE-def}). The envelope wavefunction associated with the ket $|C_\beta\rangle$ is given by Eq.~(\ref{sa-basis-def}).

For a single subband $\beta$, the wave function $\phi_\beta(z)$ and its derivative, $\phi^{'}_\beta(z)$, at coordinates $z_L$ and $z_R$ (Fig.~\ref{Fig-JScat}) are related via the following transfer matrix relation\cite{EconomouBook:2006}
\begin{widetext}
\begin{eqnarray}
\label{MC-tmx-sing}
    \left[\begin{array}{c}\phi_\beta(z_L)\\\phi_\beta^{'}(z_L)\\\end{array}\right]=
    \begin{bmatrix} 
		\cos[k^C_\beta(z_{R}-z_L)] & \sin[k^C_\beta(z_{R}-z_L)]/k^C_\beta\\
		-k^C_\beta\sin[k^C_\beta(z_{R}-z_L)] & \cos[k^C_\beta(z_{R}-z_L)]
	\end{bmatrix}    
	   \left[\begin{array}{c}\phi_\beta(z_R)\\\phi_\beta^{'}(z_R)\\\end{array}\right].
\end{eqnarray}
\end{widetext} 
Now, we introduce a column vector of the wave functions $\bm\phi(z)=[\phi_\beta(z)]$ and their derivatives $\bm\phi^{'}(z)=[\phi^{'}_\beta(z)]$ with $\beta=\overline{1,N}_{sb}$. In these notations, we recast Eq.~(\ref{MC-tmx-sing}) to the block matrix form 
\begin{eqnarray}
\label{MC-mx-def}
    \left[\begin{array}{c}\bm{\phi}(z_L)\\\bm\phi^{'}(z_L)\\\end{array}\right]=
    \begin{bmatrix} 
		\mathbf{M}^C_\text{11} & \mathbf{M}^C_\text{12} \\
		\mathbf{M}^C_\text{21} & \mathbf{M}^C_\text{22} 
	\end{bmatrix}    
	   \left[\begin{array}{c}\bm{\phi}(z_R)\\\bm\phi^{'}(z_R)\\\end{array}\right],
\end{eqnarray}
where the diagonal transfer matrix block is $\mathbf{M}_\text{11}=\mathbf{M}_\text{22}=\text{diag}[\cos[k^C_\beta(z_{R}-z_L)]]$. The upper and lower  off-diagonal blocks are  $\mathbf{M}_\text{12}=\text{diag}[\sin[k^C_\beta(z_{R}-z_L)]/k^C_\beta]$  and $\mathbf{M}_\text{21}=\text{diag}[-k^C_\beta\sin[k^C_\beta(z_{R}-z_L)]]$, respectively.   

Matching the boundary conditions for the $L$-segment wave function (Eq.~(\ref{Psi-L})) and its derivative with the $C$-segment wave function (Eq.~(\ref{Psi-C})) and its derivative, we get
\begin{eqnarray}
\label{LC-BC-eq}
    &~&\sum\limits_{\alpha=1}^{N_{sb}}\Omega^{CL}_{\beta\alpha}(\delta_{\alpha_o\alpha}e^{ik^L_{\alpha_o}z_L}+ r^L_{\alpha_o\alpha}e^{-ik^L_{\alpha}z_L})=
    \phi_\beta(z_L),
 \\\label{LC-jBC-eq}
   &~&ik^L_{\alpha}(\delta_{\alpha_o\alpha}e^{ik^L_{\alpha_o}z_L} - r_{\alpha_o\alpha}e^{-ik^L_{\alpha}z_L})= 
   	\sum\limits_{\beta=1}^{N_{sb}} \Omega^{LC}_{\alpha\beta}\phi^{'}_{\beta}(z_L).\;\;\;\;\;\;\;\;\;
\end{eqnarray}
Here, the overlap integrals $\Omega^{CL}_{\beta\alpha}$ and $\Omega^{LC}_{\alpha\beta}$ are defined in Eq.~(\ref{Sab}). Further defining a column vectors  of the Kronecker deltas, $\bm\Delta_{\alpha_o}=[\delta_{\alpha_o\alpha}]$, and a column of reflection amplitudes, $\mathbf{r}_{\alpha_o}^L=[r^L_{\alpha_o\alpha}]$, with $\alpha=\overline{1,N}_{sb}$, we recast Eqs.~(\ref{LC-BC-eq}) and (\ref{LC-jBC-eq}) to an equivalent block-matrix form that reads
\begin{eqnarray}
\label{LC-BC-mx}
\left[\begin{array}{c}\bm{\phi}(z_L)\\\bm\phi^{'}(z_L)\\\end{array}\right]&=&
    \begin{bmatrix} 
		\bm\Omega_{CL} & \mathbf{0} \\
		\mathbf{0} & \bm\Omega_{LC}^{-1} 
	\end{bmatrix}    
	\begin{bmatrix} 
		\mathbf{I} & \mathbf{I} \\
		i\mathbf{K}_L & -i\mathbf{K}_L 
	\end{bmatrix}
\\\nonumber&\times&    
    \begin{bmatrix} 
		e^{i\mathbf{K}_Lz_L} & \mathbf{0} \\
		\mathbf{0} & e^{-i\mathbf{K}_Lz_L}
	\end{bmatrix}
    \left[\begin{array}{c}\bm\Delta_{\alpha_o}\\\mathbf{r}^L_{\alpha_o}\\\end{array}\right].
\end{eqnarray}
In Eq.~(\ref{LC-BC-mx}), the overlap integral blocks are $\bm\Omega_{CL}=\bm\Omega^\dag_{LC}=\left[\langle C_\beta|L_\alpha\rangle\right]$, the unit matrix is $\textbf{I}=[\delta_{\alpha\beta}]$, the wavevector matrix is $\textbf{K}_L=\text{diag}[k^L_\alpha]$, and the exponent of the wavevector matrix is $e^{i\textbf{K}_Lz_L}=\text{diag}[e^{ik^L_\alpha z_L}]$. All $\alpha,\beta=\overline{1,N}_{sb}$. 

Similar to above, matching the boundary conditions at the $C$-$R$ interface for the wavefunctions defined in Eqs.~(\ref{Psi-C}) and (\ref{Psi-R}) and their derivatives, results in 
\begin{eqnarray}
\label{RC-BC-eq}
  &~&\phi_\beta(z_R)= \sum\limits_{\gamma=1}^{N_{sb}} \Omega^{CR}_{\beta\gamma}
    t^R_{\alpha_o\gamma}e^{ik^R_{\gamma}z_R},
 \\\label{RC-jBC-eq}
 &~&\sum\limits_{\beta=1}^{N_{sb}}\Omega^{RC}_{\gamma\beta}\phi^{'}_\beta (z_R)= 
 ik^R_{\gamma}t^R_{\alpha_o\gamma}e^{ik^R_{\gamma}z_R}.
\end{eqnarray}
This set of linear equations is equivalent to the block-matrix equation
\begin{eqnarray}
\label{RC-BC-mx}
    \left[\begin{array}{c}\bm{\phi}(z_R)\\\bm\phi^{'}(z_R)\\\end{array}\right]&=&
    \begin{bmatrix} 
		\bm\Omega_{CR} & \mathbf{O} \\
		\mathbf{O} & \bm\Omega_{RC}^{-1} 
	\end{bmatrix}    
	\begin{bmatrix} 
		\mathbf{I} & \mathbf{I} \\
		i\mathbf{K}_R & -i\mathbf{K}_R 
	\end{bmatrix}
\\\nonumber&\times&    
    \begin{bmatrix} 
		e^{i\mathbf{K}_Rz_R} & \mathbf{O} \\
		\mathbf{O} & e^{-i\mathbf{K}_Rz_R}
	\end{bmatrix}
    \left[\begin{array}{c}\mathbf{t}^R_{\alpha_o}\\\mathbf{0}\\\end{array}\right],
\end{eqnarray}
where ${\bf t}^R_{\alpha_o}=[t_{\alpha_o\gamma}]$ and $\mathbf{0}$ are the same size columns of transmission amplitudes and zeros, respectively. The overlap integral matrix is $\bm\Omega_{CR}=\bm\Omega_{RC}^\dag=\left[\langle C_\beta|R_\gamma\rangle\right]$ and ${\bf O}$ is a matrix of zeros. $\textbf{K}_L=\text{diag}[k^R_\gamma]$ and  $e^{i\textbf{K}_Rz_R}=\text{diag}[e^{ik^R_\gamma z_R}]$. All the matrices are of the $N_{sb}\times N_{sb}$ size. 

Substitution of Eqs.~(\ref{LC-BC-mx}) and (\ref{RC-BC-mx}) into Eq.~(\ref{MC-mx-def}) followed by trivial linear matrix  manipulations results in Eq.~(\ref{Scat-LR-eq}) and (\ref{Smx-LR-1}).

\section{Derivation of transfer matrix for electron emission}
\label{Apx-STnl}

As demonstrated in Sec.~\ref{Sec-Emss} electron emission does not mix electron states that belong to different subbands. Therefore, we describe this process using $2\times 2$ transfer matrix representation associated with a single subband denoted by $\alpha$. 

Let us define the electron wave function within the emission region $z_0\leq z \leq z_v$ as
\begin{eqnarray}
\label{Psi-e}
    \Psi^{e}_\alpha(z) &=& \varphi_\alpha(z)|\alpha\rangle,
\end{eqnarray}
with the ket $|\alpha\rangle$ satisfying Eq.~(\ref{sa-basis-def}). Values of the wavefunction $\varphi(z)$ and its derivative $\varphi^{'}(z)$ at the boundaries of the emission region are determined by the following equation
\begin{eqnarray}
\label{MT-tmx-1}
    \left[\begin{array}{c}\varphi_{\alpha}(z_0)\\\varphi_{\alpha}^{'}(z_0)\\\end{array}\right]=
    \begin{bmatrix} 
		{\cal M}^\alpha_{11} & {\cal M}^\alpha_{12}\\
		{\cal M}^\alpha_{21} & {\cal M}^\alpha_{22}
	\end{bmatrix}  
	   \left[\begin{array}{c}\varphi_{\alpha}(z_v)\\\varphi_{\alpha}^{'}(z_v)\\\end{array}\right].
\end{eqnarray}
with the transfer matrix defined in Eq.~(\ref{MT-tmx}).\cite{EconomouBook:2006}

Matching the boundary conditions for the electron wavefunctions given by Eqs.~(\ref{Psi-in}) and (\ref{Psi-e}) and their derivatives at $z_0$ can be performed in analogy with Eqs.~(\ref{LC-BC-eq})--(\ref{LC-BC-mx}). This results in
\begin{eqnarray}
\label{LT-BC-mx} 
    \left[\begin{array}{c}1\\ r_{\alpha}\\\end{array}\right]&=&
    \begin{bmatrix} 
		e^{-ik_{\alpha}z_0} & 0 \\
		0 & e^{ik_{\alpha}z_0}
	\end{bmatrix}  
    \begin{bmatrix} 
		1/2 & -im^*_l/(2m_ek_{\alpha}) \\
		1/2 & im^*_l/(2m_ek_{\alpha}) 
	\end{bmatrix}\;\;\;\;
\\\nonumber&\times&	
         \left[\begin{array}{c}\varphi_\alpha(z_0)\\\varphi^{'}_\alpha(z_0)\\\end{array}\right],
\end{eqnarray}
where $k_\alpha$, is given by Eq.~(\ref{kzE-def}) and the ratio $m^*_l/m_e$ account for the electron mass mismatch at the interface. 

In analogy with Eqs.~(\ref{RC-BC-eq})--(\ref{RC-BC-mx}), we match the boundary conditions at $z_v$ for the wavefunctions given by  Eqs.~(\ref{Psi-e}) and (\ref{Psi-v}) and their derivatives. This results in
\begin{eqnarray}
\label{T-BC-mx}
    \left[\begin{array}{c}\varphi_\alpha(z_v)\\\varphi^{'}_\alpha(z_v)\\\end{array}\right]=
	\begin{bmatrix} 
		1 & 1 \\
		ik^v_{\alpha} & -ik^v_{\alpha} 
	\end{bmatrix}
    \begin{bmatrix} 
		e^{ik^v_{\alpha}z_v} & 0 \\
		0 & e^{-ik^v_{\alpha}z_v}
	\end{bmatrix}
    \left[\begin{array}{c}t_{\alpha}\\0\\\end{array}\right],\;\;\;\;\;\;
    \\\nonumber
\end{eqnarray}
where  $k^v_{\alpha}$ is defined in Eq.~(\ref{kzEv-def}).

Substitution of Eqs.~(\ref{LT-BC-mx}) and (\ref{T-BC-mx}) into Eq.~(\ref{MT-tmx-1}) accompanied by linear matrix operations results in Eq.~(\ref{ST-mx-eq}) with the transfer matrix in the form of Eq.~(\ref{S-mx-eq}).

\section{Models for electron scattering by acoustic and optical phonons}
\label{Apx-Scph}

This Appendix provides a summary of the models for the electron-phonon scattering in bulk crystalline and NW structures that were implemented into the MC transport simulations discussed in Sec.~\ref{Sec:MCsmls}.  

In bulk diamond crystals, the {\em acoustic} phonons cause electron scattering {\em within} a conduction band valley with the rate
\begin{eqnarray}
\label{W3Dac}
W^\texttt{ac}(E) & = & \frac{2\pi\Xi_\texttt{ac}^2 k_\texttt{B}T}{\hbar \bar\rho v_s^2}
\\\nonumber &\times&
\frac{(2m_l^*)^{3/2}\sqrt{E(1+\alpha E)}}{4\pi^{2}\hbar^{3}}(1+2\alpha E),
\end{eqnarray}
where $E$ is the total energy of a conduction electron, $\alpha$ is the band non-parabolicity parameter, $\bar\rho$ is crystal density, $v_s$ is the sound velocity, and $\Xi_\texttt{ac}$ is the acoustic deformation potential.

The intervalley electron scattering in the bulk diamond is facilitated by the {\em optical} phonons and characterized by the following rate
\begin{eqnarray}
\label{Wopiv-bulk}
W^\texttt{op}(E)&=&\frac{\pi\Xi_\texttt{op}^2 Z_{f}}{\bar\rho\omega_{0}}\left[\bar n(\omega_{0})+1/2\mp1/2\right]
\\\nonumber&\times&
\frac{(2m_l^{*})^{3/2}\sqrt{E_{f}(1+\alpha E_{f})}}{4\pi^{2}\hbar^{3}}(1+2\alpha E_f),
\end{eqnarray}
where $E$ is the total initial energy of a conduction band electron, $E_{f}=E-\Delta E\pm\hbar\omega_{0}$ is its final energy with $\Delta E$ being the energy difference between the bottoms of initial and final valleys, $\hbar\omega_{0}$ is a quantum of phonon energy, and $Z_{f}$ is the number of equivalent final valleys. The phonons population at thermal equilibrium is $\bar n(\omega_{0})=[e^{\hbar\omega_{0}/k_\texttt{B}T}-1]^{-1}$.

In the case of NW structures, we define a set of quantum numbers $\beta_o=\{n_o,a_o\}$ designating  an electron subband (see Sec.~\ref{Sec:ScatMdls}) in which its kinetic energy is $E^z_{\beta_o}=\hbar^2k^2_{\beta_o}/2m^*_l$. The rate for this electron to scatter into subband $\beta=\{n,a\}$ due to the interaction with the {\em acoustic} phonons is\citep{Ramayya:2008} 
\begin{eqnarray}
\label{W1Dac}
W_{\beta_o\beta}^\texttt{ac}(E^z_\beta) & = & \frac{\Xi_\texttt{ac}^{2}k_\texttt{B}T}{\hbar^{2}\bar\rho v_s^2}
\\\nonumber&\times&
{\cal D}_{\beta_o\beta}\frac{(2m_l^{*})^{1/2}(1+2\alpha E^z_\beta)}{\sqrt{E^z_\beta(1+\alpha E^z_\beta)}}
\Theta(E^z_\beta).
\end{eqnarray}
Here, the electron final state kinetic energy is
\begin{eqnarray}
\label{Ebbo}
E^z_\beta=\frac{\hbar^2}{2m_t^*}\left(\kappa_{\beta_o}^2-\kappa_\beta^2\right)
	+\frac{\sqrt{1+4\alpha E^z_{\beta_o}}-1}{2\alpha}\pm\hbar\omega_0,\;\;\;\;\;\;
\end{eqnarray}
with $\omega_0=0$ set specifically for the acoustic phonons. Due to the presence of van Hove singularity in 1D density of electron states one requires that $E^z_\beta>0$ as ensured by the Heaviside step function, $\Theta(E^z_\beta)$, entering Eq.~(\ref{W1Dac}). The overlap integral in Eq.~(\ref{W1Dac}) defined in terms of the normalized Bessel functions of the first kind (Eq.~\eqref{sa-basis-def}) reads
\begin{eqnarray}
\label{Dbbo}
	{\cal D}_{\beta_o\beta}&=&N^2_{\beta_o}N^2_{\beta}
		\int\limits_0^{\rho_{NW}}d\rho \rho J^2_{n_o}(\kappa_{\beta_o}\rho)J^2_{n}(\kappa_{\beta}\rho).
\end{eqnarray}
Notice that in contrast to Eq.~(\ref{Sab}), this integral does not conserve the angular momentum quantum number and has dimensionality of the inverse area.  
 
Finally, the rate of electron in a subband $\beta_o$ of a kinetic energy $E^z_{\beta_o}$ to scatter into another subband $\beta$ and acquire a kinetic every $E^z_\beta$ (Eq.\eqref{Ebbo}) facilitated by the {\em optical} phonons is\citep{Ramayya:2008}
\begin{eqnarray}
\label{W1Dop}
W_{\beta_o\beta}^\texttt{op}(E^z_\beta)&=&\frac{\Xi_\texttt{op}^2}
{2\hbar\omega_{0}\bar\rho}\left[\bar n(\omega_{0})+1/2\mp1/2\right]
\\\nonumber&\times&
\mathcal{D}_{\beta_o\beta}\frac{(2m_l^{*})^{1/2}(1+2\alpha E^z_\beta)}{\sqrt{E^z_\beta(1+\alpha E^z_\beta)}}\Theta(E^z_\beta).
\end{eqnarray}
Here, the overlap integral, $\mathcal{D}_{\beta_o\beta}$, is also given by Eq.~\eqref{Dbbo} and $E^z_{\beta}$ by Eq.\eqref{Ebbo} with $\hbar\omega_0\neq 0$.

For the simulations in Sec.~\ref{Sec:MCsmls}, bulk diamond parameters entering the scattering rates (Eqs.~(\ref{W3Dac})--(\ref{W1Dac}), and Eq.~\eqref{W1Dop}) are adopted from Ref.~[\onlinecite{NavaSolStCom:1980}]. {\color{black} In particular, the band nonparabolicity is set to $\alpha=0$.\cite{NavaSolStCom:1980,Jacoboni_RMP:1983}}


%

\end{document}